\documentclass[useAMS,usenatbib,letterpaper]{mn2e}
\usepackage{graphicx}
\usepackage{amsfonts}
\usepackage{amssymb}
\usepackage{amsmath}
\usepackage{tabularx}
\usepackage {url}
\usepackage[]{longtable}

\title[SPIDER IX - Classifying Galaxy Groups]{SPIDER IX - Classifying Galaxy Groups according to their Velocity Distribution}

\author[A.L.B.~Ribeiro et al.]
{A.L.B. Ribeiro$^{1}$\thanks{E-mail:
albr@uesc.br}, R.R.~de~Carvalho$^{2}$, M.~Trevisan$^{2}$, H.V.~Capelato$^{2,}$$^{3}$, F.~La~Barbera$^{4}$,
\and P.A.A.~Lopes$^{5}$, A.C.~Schilling$^{1}$   \\
~~~~~~~~\\
$^{1}$ Laborat\'orio de Astrof\'{\i}sica Te\'orica e Observacional, Universidade Estadual de Santa Cruz -- 45650-000, Ilh\'eus-BA, Brazil\\
$^{2}$ Instituto Nacional de Pesquisas Espaciais, MCT, S.J. Campos, Brazil\\
$^{3}$ N\'ucleo de Astrof\'{\i}sica Te\'orica, Universidade Cruzeiro do Sul, S. Paulo,  Brazil\\
$^{4}$ INAF -- Osservatorio Astronomico di Capodimonte, Napoli, Italy\\
$^{5}$ Observat\'orio do Valongo, Universidade Federal do Rio de Janeiro, Brazil
}

\begin{document}

\date{Accepted xxx. Received xxxx}

\pagerange{\pageref{firstpage}--\pageref{lastpage}} \pubyear{2012}

\maketitle

\label{firstpage}

\begin{abstract} 

We introduce a new method to study the velocity distribution of galaxy systems, the Hellinger Distance (HD) - designed for detecting departures from a Gaussian velocity distribution. 
Testing different approaches to measure normality of a distribution, we conclude
that HD is the least vulnerable method to type I and II statistical errors. We define a relaxed galactic system as the one with unimodal velocity distribution and a normality deviation below a critical value (HD$<$0.05). In this work, we study the gaussian nature of the velocity distribution of the Berlind group sample, and of the FoF groups from the Millennium simulation. For the Berlind group sample ($z<0.1$), 67\% of the systems are classified as relaxed, while for the Millennium sample we find 63\% ($z=0$). We verify that in multimodal groups the average mass of modes in high multiplicity (${\rm N\geq 20}$) systems are significantly larger than in low multiplicity ones (${\rm N<20}$), suggesting that groups experience a mass growth at an increasing virialization rate towards $z$=0, with larger systems accreting more massive subunits. We also investigate the connection between galaxy properties ([Fe/H], Age, eClass, g-r, R$_{petro}$ and $\langle\mu_{petro}\rangle$) and the gaussianity of the velocity distribution of the groups. Bright galaxies (M$_{\rm r} \le$-20.7) residing in the inner and outer regions of groups, do not show significant differences in the listed quantities regardless if the group has a Gaussian (G) or a Non-Gaussian (NG) velocity distribution. However, the situation is significantly different when we examine the faint galaxies (-20.7$<$M$_{\rm r} \le$-17.9). In G groups, there is a remarkable difference between the galaxy properties of the inner and outer galaxy populations, testifying how the environment is affecting the galaxies. Instead, in NG groups there is no segregation between the properties of galaxies in the inner and outer regions, showing that the properties of these galaxies still reflect the physical processes prevailing in the environment where they were found earlier. 

\end{abstract}

\begin{keywords}
galaxies -- groups.
\end{keywords}

\section{Introduction}

There has been a longstanding interest in determining how galactic systems form and evolve, with many open questions though. Cosmological initial conditions determine the evolution of a galactic system, which first expands following the Hubble flow, decouples from it, reaches maximum expansion, turns around, collapses and eventually virializes (Gunn \& Gott 1972). The details of this generic scheme are still largely missing, and there is no reliable methodology for determining the dynamical and evolutionary status of a group or cluster of galaxies despite many efforts over the last thirty years (e.g. Yahil \& Vidal 1977; Menci \& Fusco-Femiano 1996; Robotham et al. 2008). In a $\Lambda$CDM scenario, more massive galaxy systems assemble their mass from the merging of less massive ones (e.g. de Lucia et al. 2006; Naab et al. 2007; Cattaneo et al. 2011) which, depending on how the merging timescale varies with mass, may imply that dynamical equilibrium also varies continuously with mass.

Observers usually collect a considerable amount of redshifts per group/cluster and measure the velocity dispersion of the system translating it into mass through the virial theorem. This process can bias the mass estimate depending how far from virial equilibrium the system is, especially for the small mass ones. Giuricin et al. (1988) suggests a correction factor to the virial mass assuming a specific model for the system's evolution, which only accounted for internal gravitational forces and neglected tidal interactions with the surroundings. In a different framework, Mamon (1993) describes the evolution of an isolated dynamical system in an expanding Universe, examining a diagram defined by the dimensionless crossing time and the dimensionless mass bias. His analysis, based on Gourgoulhon, Chamaraux, \& Fouqu\'e groups (Fouqu\'e et al. 1992, Gourgoulhon, Chamaraux, \& Fouqu\'e 1992), shows that most groups are still near the cosmological turnaround. The diagram, however, suffers from degeneracies between the expansion and early collapse phases, as well as between the full collapse and rebound phases. Also, albeit most groups follow the fundamental track, defined as the locus of virialized systems in the M/L {\it versus} crossing time diagram, there is a large scatter and the result is inconclusive (Mamon 2007). Diaferio et al. (1993), using unbiased estimators of velocity dispersion and crossing time for a sample of CfA and simulated groups, concluded that most groups are probably still in the collapse phase. However, more recently, Niemi et al. (2007) do not find any correlation between the virial ratio and the crossing time, which poses a question for using this quantity as an estimator of the stage of virialization of the systems.  

Another way of characterizing the dynamical state of a galactic system is through the study of its observed l.o.s. velocity distribution. Theoretical and phenomenological developments suggest that the virialized equilibrium state of a (spherical) gravitational systems is described, to a good approximation, by a unique and same distribution function, {\it viz}  a Maxwell-Botzmann distribution function (e.g. Ogorodnikov 1957, Lynden-Bell 1967, Hjorth \& Williams 2010, Barnes \& Williams 2012, Beraldo \& Silva 2013). In phase-space coordinates this translates to a Gaussian function (or Normal distribution). N-body numerical experiments of the relaxation of single isolated gravitational systems (Merrall \& Henriksen, 2003) or that of cosmological halos (Hansen et al, 2005, 2006) also support these conclusions. Although none of the distribution functions proposed in the literature may be considered legitimate mass-independent Maxwell-Boltzmann distribution functions, for several of them the exponential Maxwellian appears to be a very good approximation over a large range of energies, excluding the extreme values. These  limits correspond to the very center of the systems and to their outskirts, where the distribution function of secondary accretion should dominate, possibly producing the characteristic velocities anisotropy of the velocity distribution functions which are found in the outer regions of DM halos of cosmological simulations (Lemze et al, 2012, Skielboe et. al. 2012).  

It seems conceivable that by estimating the relative matching of the velocity distribution of a system to a Normal distribution, it should be possible to estimate the relaxation stage of the system towards equilibrium.  Notice that because of projection effects, the departures from the Maxwell-Boltzmann distribution function near the center of the systems, should be hardly detectable due to contamination by those galaxies that, while bound to the system, are physically far from the center.  Moreover because of the selection criteria that we will be applying to the sample of clusters galaxies, the non-Maxwell-Boltzmann effects expected to appear in the outskirts of the systems, due to secondary infall, will be automatically avoided. This suggests that discriminating groups according to their velocity distributions may be a promising way to assess the dynamics of galaxy systems. 

Beers et al. (1990) stress the difficulty in determining when a given velocity distribution differs significantly from normality, pointing out that the classification of groups as Gs, for which the observed l.o.s velocity distribution function is well fitted by a normal distribution, and NGs, may be dependent on the statistical test used. Hou et al. (2009) have examined three figures of merit (Anderson-Darling, Kolmogorov and $\chi^2$ tests) to find which statistical tool separates better between G and NG galaxy groups. Using Monte Carlo simulations and a sample of groups selected from CNOC2, they found the Anderson-Darling test to be far more reliable at detecting real departures from normality. Also, G and NG groups exhibit distinct velocity dispersion profiles, suggesting they are in different dynamical stages. About 68\% of the CNOC2 groups are found to  be G. Hence, the choice of the statistical test to be applied on data is crucially important in the subsequent analysis of galaxy groups. However, it is hard to establish how trustful the statistical methods to test normality are. Sample size effect is a potential problem for all the hypothesis tests presented in the literature. 

In the present work, we introduce a methodology to quantify deviations from normality, using a stable approximation to the Fisher information metric -- the so called Hellinger distance (hereafter HD). The paper is organized as follows: Section 2 gives a description of the SDSS and Millennium galaxy group samples. Section 3 presents the HD methodology, describes how we have calibrated it against the sample size bias, presents a comparison between HD with some normality/unimodality tests, and
sets robust criteria to identify G systems with respect to their velocity distributions. Section 4 presents our results for the SDSS and Millennium samples. Section 5 has a study of the galaxy properties in G and NG systems. Finally, in Section 6 we discuss our main findings. 

\section{Data}
\subsection{The FoF group catalog}
\label{sec:group_catalog}

The updated FoF group catalogue from the SDSS-DR7 contains $10,124$ systems 
and the method is described in Berlind et al. (2006). This new version differs 
from the first one only in the larger area used ($9,380$ square degrees, 
compared to the original $3,495$ square degrees from DR3). In addition to the 
coordinates and central redshift, the list provides estimates of the total 
number of galaxies in the group and its velocity dispersion. However, we
only consider the positional and redshift information, 
re-deriving the member list and
group properties (velocity dispersion, radius and mass). When deriving group 
properties we adopt the procedure described in La Barbera et al. (2010), with
a few modifications to avoid defining new structures around the original 
systems selected by the FoF algorithm. As in La Barbera et al. (2010) we 
derive a new central redshift for the groups, applying the gap technique 
(Adami et al. 1998), Lopes 2007, Lopes et al. 2009a) to the central 
(0.67 Mpc) galaxies. This technique separates groups after identifying
gaps in the redshift distribution larger than a given value. We are able 
to re-derive redshifts for $8,635$ ($85.3$\%) systems of the $``original$'' 
sample (note that we require at least three galaxies in this analysis). 
As explained below those new redshifts are not used. We applied the gap 
technique only to remove very poor groups, with less than three galaxies.

For these groups we apply the ``shifting gapper'' technique 
(Fadda et al. 1996) as in Lopes et al. (2009a) in order to obtain a list 
of group members independent of the FoF algorithm. This method consists
of applying the gap-technique in radial bins from the cluster centre. The bin 
size is 0.56 Mpc or larger to force the selection of at least 15 galaxies 
(consistent with Fadda et al. 1996). Galaxies not linked to the
main body of the cluster are eliminated.  We obtain an estimate of the 
line-of-sight velocity dispersion and then perform a virial analysis,
analogous to the one described in Girardi et al. (1998), 
Popesso et al. (2005, 2007), Biviano et al. (2006), and Lopes et al. (2009a). 
First, the projected 'virial radius' ($R_{PV}$) is derived and a first 
estimate of the virial mass is obtained (using equation 5 of 
Girardi et al. 1998). The surface pressure correction is applied to the 
mass estimate and a Navarro, Frenk \& White (1997, hereafter NFW) profile 
is assumed to obtain estimates of $R_{500}$, $R_{200}$, $M_{500}$ and $M_{200}$. 
These parameters are derived for most groups from 
the FoF sample. When performing the virial analysis we require 
at least five galaxies within the maximum radius considered 
($R_{max} =$ 4.0 Mpc).\footnote{In this work, cosmology is defined by $\Omega_m$ = 0.3, $\Omega_\lambda$ = 0.7, and $H_0 = 100~h~{\rm km~s^{-1}Mpc^{-1}}$. Distance-dependent quantities are calculated using $h=0.75$.} Of the 8,635 groups we 
are able to measure the above parameters for $5,352$ (those with the 
minimum of five galaxies), i.e. $52.8$\% of the 
initial full sample. To avoid issues with groups having very close systems or 
that may be severely affected by the large-scale structure, we use the 
original coordinates and redshift from the FoF group catalog when selecting 
members and excluding interlopers. The final sample we considered
contains $5,352$ groups from the updated FoF group catalogue, with redshift $z_{max} = 0.106$ and $N_{min}=5$.
The sample was originally selected to have a uniform magnitude limit M$_r=-20$. However,
the survey magnitude limit varies with redshift.
The sample is complete for M$_r \leq {\rm M}^\ast +1$ (M$^\ast =-21.42$), with a subsample of galaxies with
 M$_r \leq {\rm M}^\ast +4$ from objects at $z\leq 0.03$.  This subsample is used in
Section 5 to analyse the properties of galaxies in G and NG groups.

\begin{figure} 
\begin{center}
\includegraphics[scale=0.45]{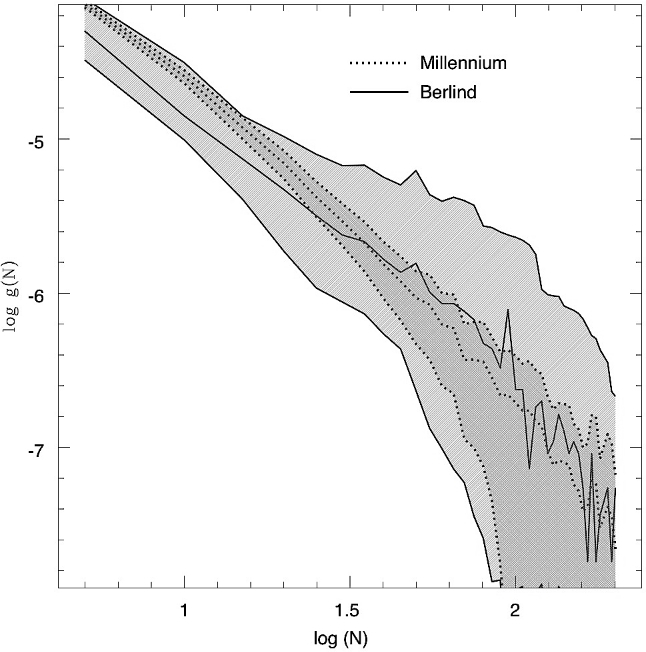}
\caption{Multiplicity function of galaxy groups -- number of group members per Mpc$^{3}$. 
The solid and dashed lines depict the Berlind's catalog
and the Millennium simulation, respectively. The shaded areas  indicate the bootstrap errors computed
as the dispersion over 1,000 bootstrap realizations of each sample.}
\end{center}
\label{}
\end{figure}

\subsection{The Millennium Simulation Catalog}

We use the dark matter Millennium Simulation, performed by the Virgo Consortium (Springel et al. 2005), and the semi-analytical galaxy formation model of de Lucia \& Blaizot (2007). The Millennium Simulation follows the evolution of a comoving cube of size 500 Mpc $h^{-1}$ in a $\Lambda$CDM cosmology, with $2 \times 10^{6}$ particles and a mass resolution of $8.6 \times 10^8 h^{-1} M_{\odot}$. As described in Springel et al. (2005), a group catalog for the Millennium Simulation is created using a FoF algorithm. Substructures (i.e. sub-halos) in a given group are then identified through the SUBFIND algorithm (Springel et al. 2001). The largest sub-halo of an FoF group typically contains more than 90$\%$ of total group mass, and is referred to as the {\it main group halo}. The semi-analytical model of de Lucia \& Blaizot (2007) is a slightly modified version of that presented by Croton et al. (2006), which simulates the evolution of galaxies within the dark matter Millennium Simulation, 
including {\it ad-hoc} recipes to describe different physical processes, like galaxy mergers, SN feedback, and suppression of cooling flows by ``radio-mode'' AGN feedback.  The semi-analytical model is accessible online through the Virgo-Millennium Database~\footnote{$http://www.g-vo.org$} (Lemson et al. 2006). For each model galaxy, the Database provides magnitudes in a variety of passbands, including SDSS (AB) ugriz. Magnitudes are computed with the Bruzual \& Charlot (2003) stellar population synthesis code, assuming a Chabrier (2003) IMF.  We queried the Millennium Database for all galaxies residing in FoF group identified in the $z=0$ simulation snapshot, with group mass, $M_{200}$, larger than $10^{13}M_{\odot}$, and galaxy magnitude brighter than $-19$ in the SDSS $r$-band.  Both these selections match closely those performed on the data. The group mass threshold is similar to the lowest $M_{200}$ measured through the virial analysis from the Berlind group catalog.  The Millenium catalog
is selected down to M$_r = -19$, i.e. 1~magnitude fainter than the Berlind cutoff. We decided
to keep a deeper cutoff in order to have better statistics in each group (especially at
low mass). Also, one should notice that there is no straightforward way to compare magnitudes
from data and simulations, as they are intrinsically defined in  different ways. On the other hand, 
we also verified that the results presented in Section 4 
remain approximately unchanged when cutting the Millenium catalog at M$_r=-20$ (rather than M$_r=-19$). 

There are $70,259$ Millennium groups with mass larger than $10^{13}M_{\odot}$.  
The query yields $883,201$ model galaxies residing in these groups, with position 
and velocity components of each galaxy being provided $wrt$ to those of the parent galaxy group. 
In Figure 1, we compare
the number density of groups in the Berlind's catalog and in the
Millennium simulation as a function of the number of group members, N -- that is, 
the group multiplicity function represented by g(N). Bootstrap errors were computed as 
the dispersion over 1,000 bootstrap realizations of the Berlind's sample. Note that both
samples have similar behaviour within these errors, allowing us to directly compare them
in the subsequent analysis of this work.

It is important to note that the Millennium simulation is used here as a benchmark to help understanding the results we find for the Berlind's catalog. The difference in number of systems in both catalogs reflects the difference in comoving volumes. The ratio of the number of groups in Millennium by the number of groups in Berlind's, with masses greater than $10^{13.75}M_{\odot}$, is $\sim$6.3, while the ratio of the comoving volumes is 6.9.


\section{Methodology}

\subsection{Hellinger Distance}

In this work, we introduce a new estimator of the distance between the empirical velocity distribution of galaxies in a group and the theoretically expected Gaussian distribution function.  The concept of distance between empirical and theoretical distributions arises naturally in probability theory, seen as an applied branch of the measure theory (see Kolmogorov 1933). There have been many metrics designed to quantify the distance between probability density functions (see Kass \& Vos, 1997 and Amari \& Nagaoka, 2000). Here, we adopt the so-called Hellinger distance, a stable approximation to the Fisher information metric (see e.g. Amari 1985).  For a discrete space,

\begin{equation}
HD^2(p,q) = 2 \sum_{x\in X} [\sqrt{p(x)} -\sqrt{q(x)}]^2,
\end{equation}

\noindent where $p$ and $q$ are probability distributions, and $x$ is a random variable.
The possible values of HD are in the range $[0,\sqrt{2}]$, but in this work we follow LeCam (1986) 
and normalize its range of possible values to $[0,1]$. The codes we have used to determine the Hellinger distance are publicly available in the R language and environment (R Development Core Team) under the distrEx (Ruckdeschel et al. 2006). R is an open-source free statistical environment developed under the GNU GPL (Ihaka \& Gentleman 1996, http://www.r-project.org). 

\subsection{HD Calibration}

Before estimating HD using real data, we have to calibrate our distance estimator to be independent of the number of objects in the system. Basically, the issue arises from the fact that the sample size usually bias the results. We can easily test the size effect on probability distances by computing HD for realizations of normal distributions as a function of the sample size, N. We perform 1,000 realizations for each N. The result is presented in Figure 2 (see the red curve above the dashed line). The expected value for HD for all these normal distributions is zero.  However, the sample size effect produces a strong 
and systematic bias. We measure the effect by fitting the biased curve with a cubic spline function. Then, later on when we measure the HD for a given group of size N, we subtract the correction 
here determined (i.e. the value of the fitted curve at N)
in order to remove the size effect. Gaussianity is set by a $3\sigma$ threshold (HD$<0.05$). 

\begin{figure} 
\begin{center}
\includegraphics[scale=0.48]{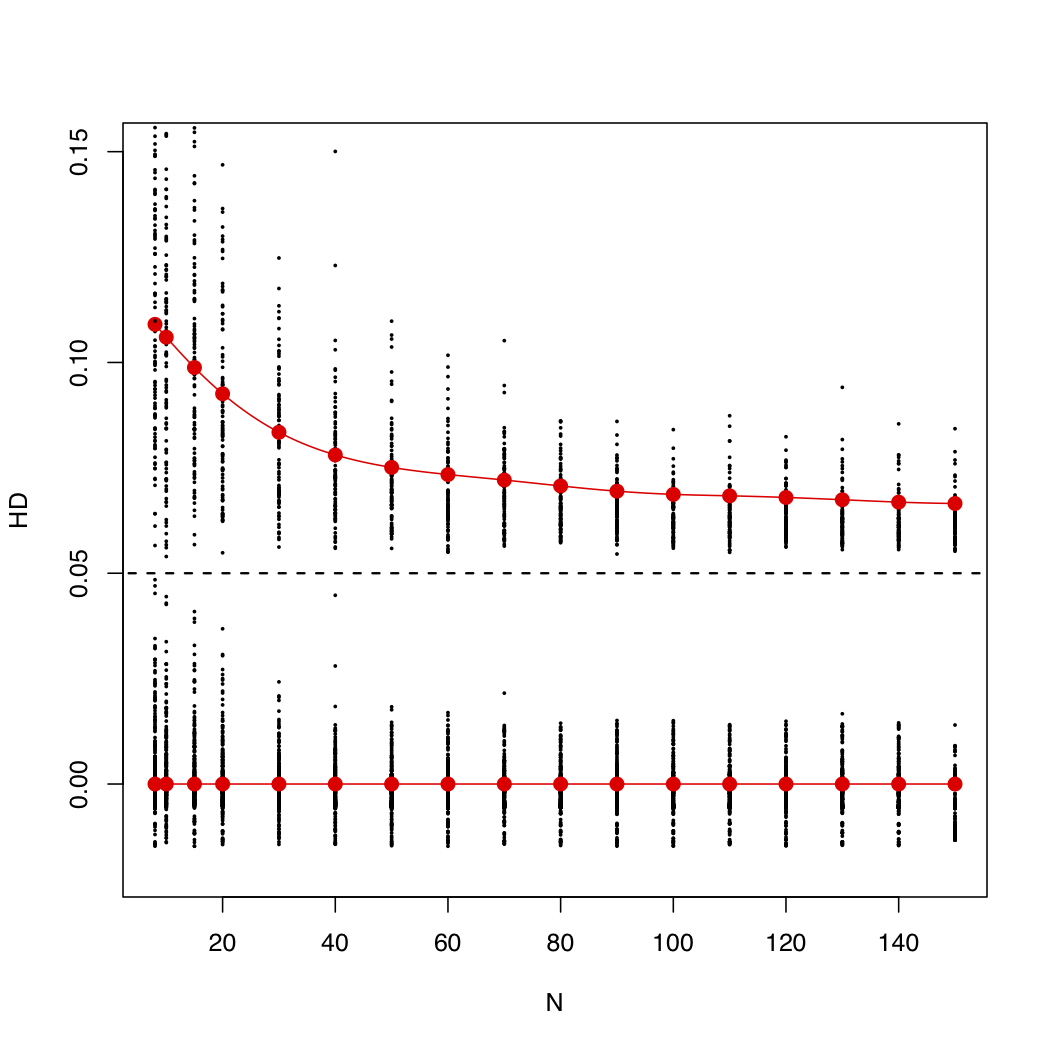}
\caption{HD computed for different sample sizes. The red solid line indicates the cubic spline fit to the data. Red points above (and below) the black dashed line indicate the median values of HD for each set of 1,000 normal realizations with size N before (and after) the correction given by the fit. The dashed line indicates the $3\sigma$ threshold below which we define a G system.
}
\end{center}
\label{}
\end{figure}

\subsection{Comparison of Methods}

The statistical power is the ability of a test to detect a desired effect -- a true normality deviation in the present case. Normality tests are based on different characteristics of the normal distribution and the power of these tests varies depending on the nature of the non-normality and the sample size (Seier, 2011). These effects can lead to confusing results. For instance, consider the following tests: 1) the Shapiro-Wilk (SW) has more power than the $\chi^2$ and the Kolmogorov-Smirnov against the symmetric short-tailed, asymmetric and both short-tailed \& long-tailed alternatives (see Shapiro \& Wilk 1965); 2) SW has more power than the Jarque-Bera test (JB) for distributions with short tails, especially if the shape is bimodal (Thadewald \& B\"uning, 2004); and 3) JB has more power than the SW test against the alternatives from the Pearson-family distributions (Jarque \& Bera, 1987). 

Hou et al. (2009) have examined three methods (Anderson-Darling, Kolmogorov and $\chi^2$ tests) to find which statistical tool distinguishes better between relaxed and non-relaxed galaxy groups. Using Monte Carlo simulations and a sample of groups selected from CNOC2, they found the Anderson-Darling test to be more reliable at detecting significant departures from normality. However, comparison of tests can lead to ambiguous results as they depend on the specific normality deviation imprinted in the data (e.g. Seier, 2011).  No test can be qualified as the most powerful against all non-normalities at the same time. Likewise, for any test of normality, it is important to determine its sensitivity to the sample size. In this respect, we perform Monte Carlo simulations of different non-normal distributions taken from the $\alpha$-stable distribution. Stable functions are a class of probability distributions allowing a wide range of skewness and kurtosis. Except in certain specific cases, the density function of an $\alpha$-stable random variable cannot be given in closed form. However, its characteristic function can always be given.

{\small
\begin{equation}
\phi(x)=
\begin{cases}
\exp{\{ix\delta - \gamma^\alpha|x|^\alpha
[1-i\beta\,{\rm sgn}(x)\tan({\pi\alpha\over2})]\}}, & \mbox{if } \alpha~\neq~1\\
\exp{\{ix\delta - \gamma|x|[1+i\beta\,{\rm sgn}(x) {2\over\pi}\ln{|x|}\}}, & \mbox{if } \alpha=1
\end{cases}
\end{equation}}

~~~~~~

\subsection{The $\alpha$-function parameters}

The $\alpha$-stable function allows a wide range of skewness and kurtosis, according to the choice of four parameters (Nolan 1998). 

\begin{enumerate}
\item[$\alpha$ -] The stability parameter, which describes the weight of the tails of the distribution:
$0 < \alpha \leq 2$. The smaller the $\alpha$ the heavier the tails. Heavy-tailed distributions are probability distributions whose tails are not exponentially bounded, that is, they have heavier tails than the exponential distribution.

\item[$\beta$ -] The symmetry parameter, controls the skewness of the distribution: -$ 1\leq \beta \leq 1$. If $\beta$ = 0 the distribution is symmetric otherwise is skewed.

\item[$\gamma$ -] The kurtosis parameter is similar to the variance of a normal distribution: $\gamma > 0$.

\item[$\delta$ -] The location parameter: $-\infty < \delta < \infty$.

\end{enumerate}

When $\alpha = 2$, the $\alpha$-stable distribution becomes a normal one.  In this case the $\beta$ parameter  is redundant, with $\gamma$ and $\delta$ corresponding to the variance and mean of the distribution (see e.g. Nolan, 1998). To compare the power of the different tests, we run realizations of $\alpha-$stable distributions with $\alpha \in (0,2)$, $\beta \in (-1,1)$ (excluding zero), $\gamma \in (0,2]$, and $\delta=0$. We also run realizations of multimodal systems, assumed here as Gaussian mixtures of 2 and 3 components. Multimodal systems are NG by definition and can be a frequent cause of normality deviation (see e.g. Ribeiro et al. 2011). In addition to the tests already mentioned, we have included the Dip test to study multimodal samples (see a description of the test and how to generate multimodal samples in Section 3.6).

\subsection{The algorithm}

First, to find type I errors (those that occur when the null hypothesis is true but rejected), data are generated as follows:

\begin{enumerate}
\item	 Generate 1000 data sets, each with size N, from a normal distribution ($\alpha=2$ in $\alpha$-stable distribution).
\item	 For each replication in Step 1, perform these normality tests: Anderson-Darling, Shapiro-Wilk, Jarque-Bera (the robust version), and Dip test. Also, compute the HD.
\item	 The error count is given by the number count$_{[{\rm test}]}$.
For each test, if p-value $\leq$ 0.05, 
count$_{[{\rm test}]}$ is increased by one (count$_{[{\rm test}]}$=
count$_{[{\rm test}]}$ + 1). 
\item   Similarly, if HD $\geq$ 0.05, the number count$_{[{\rm HD}]}$ is increased by one 
(count$_{[{\rm HD}]}$= count$_{[{\rm HD}]}$ + 1).
\item	 Determine the average Type I error rates by dividing the 
final counts by 1,000 for each normality test and HD measurement.
\item	 Repeat Steps 1 to 5 for sample size N between 8 and 150.
\end{enumerate}

Then, to find the type II errors (those that occur when the null hypothesis is false and erroneously taken as true), we replace step [1] by realizations of non-normal distributions ($\alpha\ne0$) plus multimodal cases (including the Dip test), and steps [3] and [4] to have p-value $>$ 0.05 and HD $<$ 0.05.

The results are summarized in Figure 3. As expected, the HD method produces almost no type I error, after the calibration described in Section 3.2. As a consequence, it is the best method for any sample size. At the same time, the HD measure does not introduce severe type II errors for N$>$20 in the case of skewed samples and for N$>$10 for kurtotic samples, being the best choice in comparison with all statistical tests considered here. Hence, we conclude that the HD is a reliable discriminator between G and NG systems for unimodal distributions. However, for multimodal samples, we see that the Dip test is the best choice for N$<$70, being almost equivalent to AD and SW tests for larger samples. This suggests we should pay more attention to Gaussian deviations due to multimodality.

\begin{figure}  
\begin{center}
\includegraphics[scale=0.5]{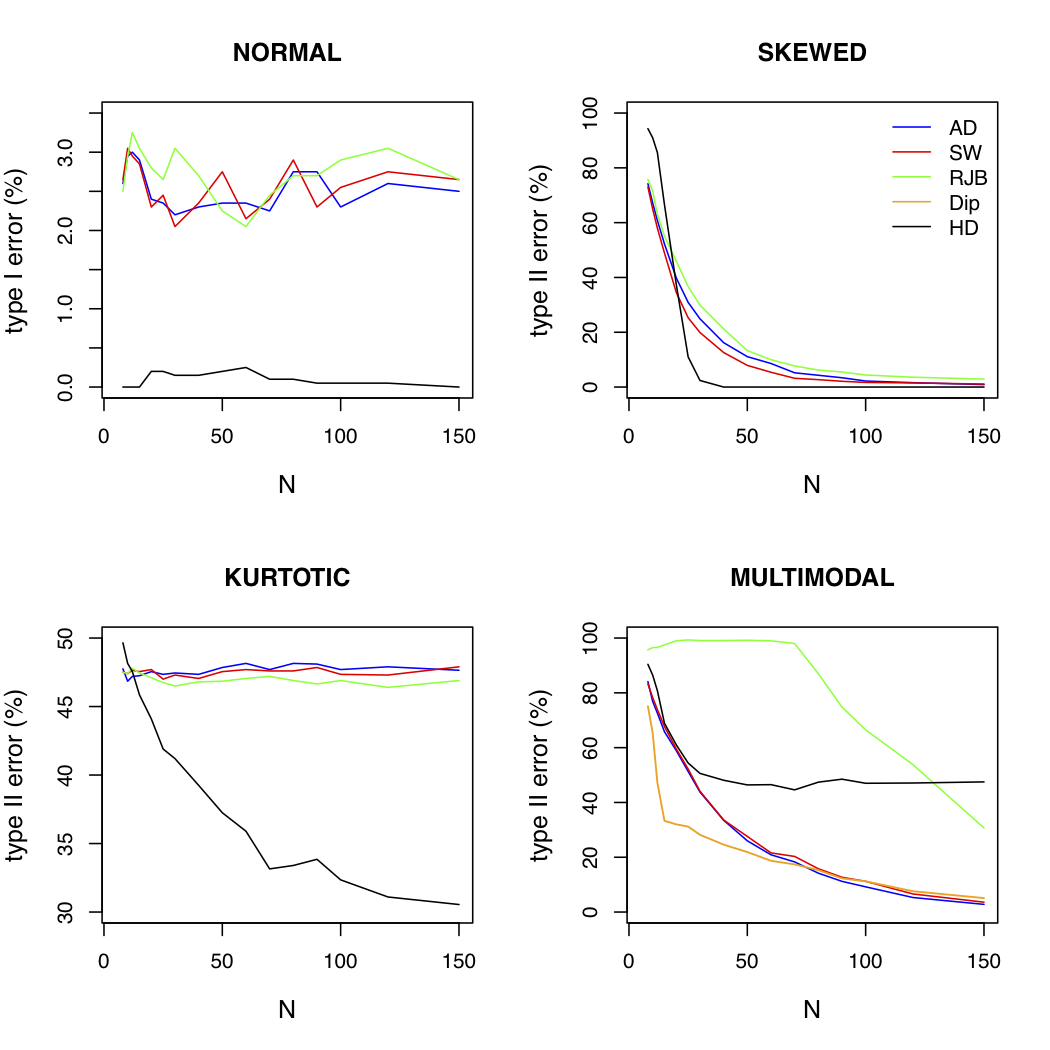}
\caption{Comparison of tests for type I and II errors, for different cases of Gaussian deviations.
Colors follow the legend in the upper right panel: HD (black), AD (blue), SW (red), RJB (green), and Dip (orange).}
\end{center}
\label{}
\end{figure}

\subsection{Multimodality}

Recent works indicate that multimodal velocity distributions may be very common in galaxy systems (e.g. Ribeiro et al. 2011, Hou et al. 2012, Einasto et al. 2012a). Thus, we also need to choose an efficient discriminator between unimodal and multimodal groups.  At first, one could imagine that multimodal objects should be classified as NG by any normality test or the HD. Multimodality depends on the separation and widths of the modes (see Ashman et al. 1994), such that a particular aggregate can be wrongly accepted as a single mode by some statistical tools. In this work, we compare the following multimodality discriminators: Dip test, Mclust method, and the Dirichlet process, briefly described here.

\begin{enumerate}

\item[(a)] The Dip test is based on the cumulative distribution of the variable of interest (Hartigan \& Hartigan 1985). The Dip statistic is the maximum distance between the cumulative input distribution and the best-fitting unimodal distribution. In some sense, this test is similar to the Kolmogorov-Smirnov test but the Dip test searches specifically for a flat step in the cumulative distribution function, which corresponds to a ``$dip$" in the histogram representation. The Dip test has the benefit of being insensitive to the assumption of Gaussianity and is therefore a true test of modality (e.g. Pinkney et al 1996; Muratov \& Gnedin 2010).

\item[(b)] Mclust is a contributed R package for normal mixture modeling and model-based clustering. 
It provides functions for parameter estimation via the Expectation-Maximization (EM) algorithm for normal mixture models with a variety of covariance structures (Fraley \& Raftery 2007). The method is based on a search of an optimal model for the clustering of the data among models with varying shape, orientation and volume. It finds the optimal number of components and the corresponding classification (the membership of each component). Mclust calculates for every galaxy the probabilities to belong to every component. The uncertainty of the classification is defined as one minus the highest probability of a galaxy to belong to a component. The mean uncertainty for the full sample is used as a statistical estimate of the reliability of the results. Recently, Einasto et al. (2012a,b) used this method to find multimodality in rich clusters they identified on SDSS-DR8 data using a FoF algorithm.

\item[(c)] The Dirichlet process is a method suitable for investigating mixed distributions with $k$ unknown components.  In the  Dirichlet Process Mixture (DPM) model, each galaxy is assumed to come from one of the components in the mixture, but we do not know which. The algorithm uses a parameter which specifies which component each galaxy came from -- what characterizes a specific mixture or a ``model". The DPM methodology relies on running Markov Chain Monte Carlo (MCMC) simulations for exploring different models (see Diebolt \& Robert 1994). This sampling process is initialized by taking models at random. Then, the code iteratively assign galaxies to the different models. The quality of the fit between the data and each model is expressed as the probability that the galaxy velocity distribution was generated by that model. By examining how often galaxies are assigned to the same model, we can get information about the best model supported by data. In this work, we assume that each component in the mixture is a Gaussian, and we find $k$ using the R language and environment (R Development Core Team) under the dpmixsim library (da Silva 2009). Ribeiro et al. (2011) used this method to study multimodality in 2dF groups.

\end{enumerate}

To compare the power of the different methods, we run Monte Carlo simulations to generate three datasets: unimodal (a single Gaussian distribution), bimodal and trimodal systems; these latter correspond to Gaussian mixtures with modal fraction $p_k$, mean $\mu_k$, standard deviation $\sigma_k$ and pairwise separation $D_{ij}=|\mu_i - \mu_j|/[(\sigma_i^2 + \sigma_j^2)/2]^{1/2}$. According to Ashman (1994), $D_{ij}>2$ is required for a clean separation between the modes. We run Monte Carlo simulations with $p_k$ in the range $(0,1)$ ($\sum {p_k} =1$), $\mu_k$ in $(-0.5,0.5)$ $(\mu_i \ne \mu_j)$, $\sigma_k$ in $(0.5,1.5)$, and $D_{ij}$ in $(1,3)$.

The results are presented in Table 1. For type I errors, we see that the Dip test is the best choice for any sample size, while for type II errors, Mclust has a slightly better performance for N$\geq 20$. Note that for type II errors we have included the AD test and the HD in the comparison, since both should reject Gaussianity for multimodal systems. Actually, while the HD and Mclust could be good choices for bimodal systems with N$> 20$, both HD and AD tests are not good choices for trimodal systems with N$\leq 50$. All these findings suggest that for small size samples (N$\leq 20$) we should use the Dip test, and for larger samples we should apply the Mclust method to identify multimodal systems.

\begin{center}
\begin{table}
\caption{Comparison of unimodality indicators: the Dip test, the Mclust method, and the DPM process.
Scores of Type I and Type II erros for controlled samples with different sizes.}           
\label{tab1}   
\begin{tabular}{l c c c}       
\hline  
               &                Type I Errors & (unimodal)     &                 \\
\hline
 N           & Dip & Mclust & DPM \\ 
\hline 
10       & 2\%   & 18\%  &  36\%    \\    
15       & 2\%  &  18\%  &  11\%    \\ 
20      &  2\%  &   14\% &  8\%    \\ 
30      &  0\%  &   7\% &  8\%    \\
40       &  0\%  &  6\%  &  8\%    \\
50 	&  0\%  &  2\%  & 4\%       \\
80	 &  0\%  &  2\%  & 4 \%     \\
100	&  0\%  &  2\%  &  3\%    \\
\hline
\end{tabular}
\begin{tabular}{l c c c c c}
               &              &     Type II Errors  & (bimodal) &   &\\

\hline
N          &  AD & Dip & Mclust & DPM & HD\\ 
\hline
10      &    37\% & 35\%   &  39\%  &  37\%  &  30\%  \\    
15      &    19\% & 19\%  &   21\%  &  19\%  &  17\% \\ 
20      &    17\% &  7\%  &   8\% &   7\%  &  13\% \\ 
30      &    12\% &  5\%  &   4\% &   8\% &   3\% \\
40      &   11\%   &  4\%  &  3\%  &  4\%   &   2\% \\
50 	&   10\% &  4\%  &  2\%  & 4\%   &   2\% \\
80	&   9\%  &  3\%  &  2\%  &  3\%  &   2\% \\
100	&   7\%  &  2\%  &  2\%  &  3\%  &   2\% \\
\hline
              &              &      Type II Errors & (trimodal)  &   &\\

\hline
N          &  AD & Dip & Mclust & DPM & HD\\ 
\hline
10      &    45\% & 42\%   &  39\%  &  38\%  &  39\%  \\    
15      &    29\% & 23\%  &   22\%  &  27\%  &  27\% \\ 
20      &    17\% &  16\%  &   14\% &   17\%  &  15\% \\ 
30      &    15\% &  12\%  &   7\% &   16\% &   13\% \\
40      &   15\%   &  9\%  &  4\%  &  9\%   &   12\% \\
50 	&   12\% &  5\%  &  4\%  & 7\%   &   10\% \\
80	&   11\%  &  4\%  &  3\%  &  4\%  &   5\% \\
100	&   7\%  &  4\%  &  3\%  &  3\%  &   4\% \\
\hline
\end{tabular}
\end{table}
\end{center}

\section{Analysis}

The tests performed in the previous section allow us to set the following two criteria to define a G group according to its velocity distribution:

\begin{enumerate}
\item[$\mathcal{C}$1:]  Admit only small deviations from the normal distribution. We impose that HD$<0.05$ for these groups.
\item[$\mathcal{C}$2:]  Must be unimodal. We thus impose that for groups with N$\leq 20$, $p>0.05$ for the Dip test. For groups with N$> 20$,  $n_{modes}=1$ for the Mclust clustering technique. 
\end{enumerate}

Groups that do not follow these criteria are considered NG. Recalling the discussion on the equilibrium distribution function of gravitational systems (Section 1), we expect  G groups to have already attained equilibrium through collapse and virialization. 

\subsection{Results for the Berlind's sample} 

Applying the Gaussianity criteria to the Berlind's sample (described in Section 2.1), we find that 67\% of them are G. For N$\leq 20$, the fraction of G systems is 66\%, and for N$> 20$, this fraction is a little higher, 70\%. An important question is whether the virial masses significantly differ between G and NG groups. Following Ribeiro et al. (2011), we have applied the Cram\'er-von Mises two-samples test on 1,000 bootstrap replicas of mass distribution for G and NG groups separately, which resulted that G and NG groups have consistent mass distributions, with p-value=0.68. For groups with N$\leq$20 we still find consistency between the G and NG mass distributions, p-value=0.12. However, for groups with N$>$20 we  find an inconsistency between the mass distributions of G and NG systems: p-value$< 10^{-5}$.  NG groups have average mass $\approx$1.4 times larger than that of G groups with N$>$20. This is consistent with the previous results of Ribeiro et al. (2011) and Krause et al. (2013), who find that NG systems 
are estimated to be larger and more massive
than the G ones, a possible consequence of applying the virial analysis to non-virialized 
groups.

\begin{figure} 
\begin{center}
\includegraphics[scale=0.5]{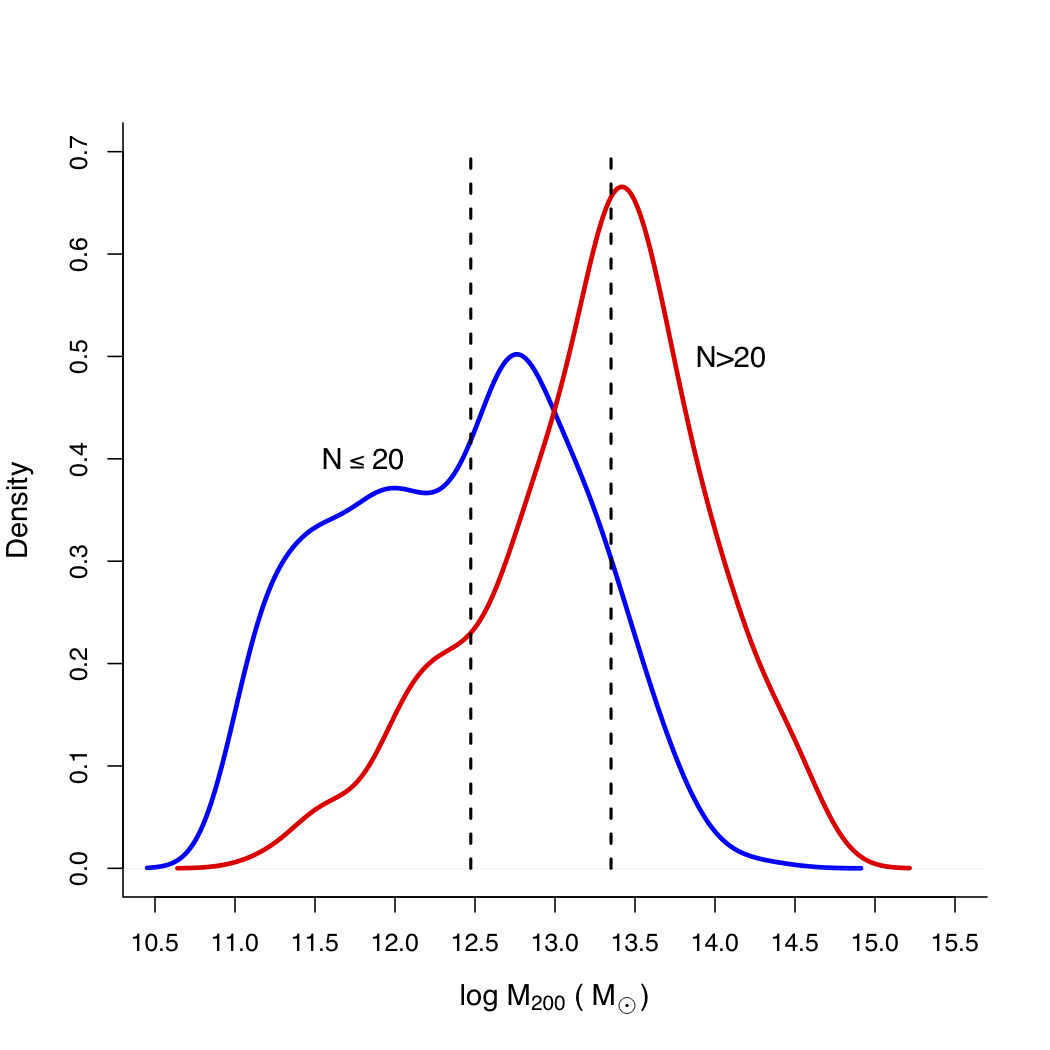}
\caption{Mass distributions in modes for systems with $N\leq20$ (blue) and $N> 20$ (red). Dashed lines mark the median for each distribution.}
\end{center}
\end{figure}

\begin{table}
\caption{Fraction of groups with substructures according to the $\beta$ test, the $\Delta$ test, and the Mclust 
method in 2 or 3 dimensions (${\rm MC_{2D}}$ and ${\rm MC_{3D}}$).}           
\label{tab1}   
\begin{tabular}{l  c c c c c c}       
\hline
 Method  & ${\rm G_{all}}$ & ${\rm NG_{all}}$ & ${\rm G_{N\leq20}}$ & ${\rm NG_{N\leq20}}$ &  ${\rm G_{N> 20}}$ & ${\rm NG_{N> 20}}$\\ 
\hline 
${\rm \beta}$   & 17\% & 17\%    & 13\%   & 6\%  &  27\% & 32\%   \\    
${\rm \Delta}$  & 12\% & 16\%    & 11\%  &  10\%  &  17\% & 34\%   \\ 
${\rm MC_{2D}}$ & 22\% & 20\%    & 29\%  &  19\%  &  39\% & 43\%   \\
${\rm MC_{3D}}$ & 23\% & 21\%    & 19\%  &  15\%  &  33\% & 43\%   \\
\hline
\end{tabular}
\end{table}

\subsubsection{Mass distribution in the subunits detected in the multimodality analysis}

Multimodality may be exhibiting the hierarchical process in action and this is a crucial point for comprehending the relation between environment and galaxy properties. Studying multimodal systems with Mclust, and assuming that all modes with at least 3 galaxies are physical entities, we estimate their virial masses according to Carlberg et al. (1997). We find that the mass distribution of modes in groups with $N\leq20$ and $N> 20$ are quite different as displayed in Figure 4 (p-value $<10^{-5}$ for the Cram\'er-von Mises and Kolmogorov-Smirnov tests), with many modes having typical masses of galaxies in small groups ($\sim$ 43\% of them have $M\leq 10^{12}~M_{\sun}$, while only $\sim$5\% of modes have masses below this value in groups with $N>20$). We also find that $\langle M_{N>20}^{modes}\rangle \simeq (14\pm 2) \langle M_{N\leq 20}^{modes}\rangle$. This result reveals that multimodality probably reflects the hierarchy of structure formation: in small systems, modes correspond to aggregates of low mass infalling into the groups, while in larger systems they correspond to larger aggregates already on the way to virialization. An important caveat is that multimodality/non-Gaussianity in velocity distributions does not necessarily mean we are detecting substructures in 2D or 3D. For instance,
Einasto et al. (2010) showed in their study of the richest galaxy clusters in the Sloan Great Wall that in several regular clusters galaxy velocity distributions have multiple components.

To investigate this connection we computed the fraction of groups with substructures according to the following methods: $\beta$ test (West et al. 1988), $\Delta$ test (Dressler \& Schectman 1988), and Mclust in 2 or 3 dimensions (${\rm MC_{2D}}$ and ${\rm MC_{3D}}$). The results are summarized in Table 2, where we see that: (i) in low multiplicity systems ($N\leq20$), the fraction of systems with substructures (6-19\%, depending on the substructure test used) is lower than the fraction of NG systems (34\%) implying a significant percentage of groups with NG velocity distribution but no substructure;  (ii) for low multiplicity systems, no method indicates that NG groups have higher fractions of substructures than G groups; (iii) high multiplicity groups have systematically higher fractions of objects with substructures in NG systems, namely the association between the NG character of the velocity distribution and the degree of substructure is only observed in the $N> 20$ systems, an important fact when interpreting substructure
as the dynamical youth of a galactic system.

\begin{figure} 
\begin{center}
\includegraphics[scale=0.5]{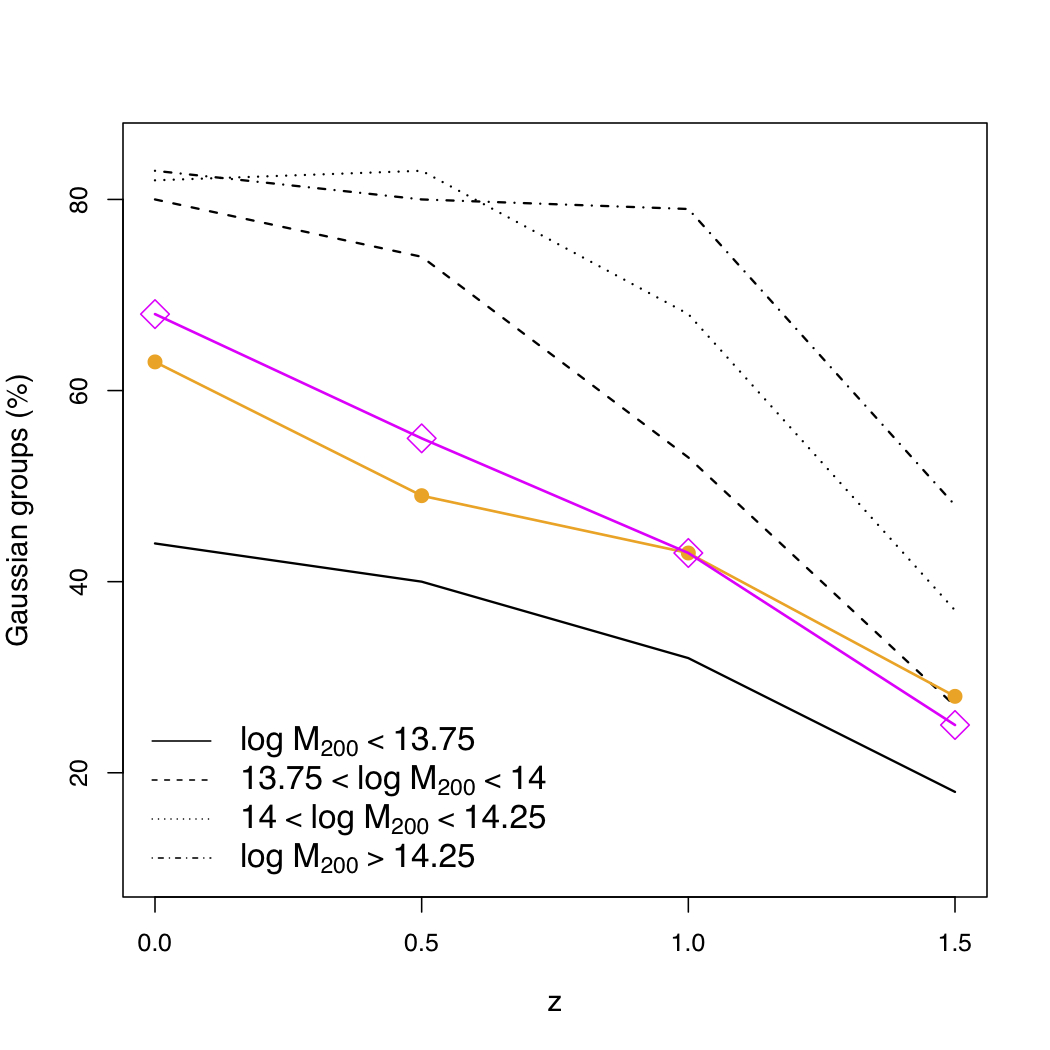}
\caption{Evolution of G groups with redshift by mass interval for the Millennium simulation data.
The orange and magenta lines (full circles and open diamonds) 
indicate the evolution for systems with all masses with limiting
magnitudes M$_r=-19$ and M$_r=-20$, respectively).
Legends with mass ranges are at the bottom left corner of the figure.}
\end{center}
\label{}
\end{figure}

\subsection{Results for the Millennium group/cluster sample}

The multimodality analysis indicates that when we examine low and high multiplicity systems we may be witnessing mass assembly in the structures and its way to dynamical equilibrium. Therefore, a
further topic of interest is the estimate of the rate at which galaxy systems achieve equilibrium. To explore this point, we have used the Millennium simulation data (described in Section 2.2). Galaxies in clusters from Millennium were selected in the same way as we do for Berlind's (described in Section 2.1). We assume here that Gaussianity + Unimodality is an indicator of virialization. Thus, following the same criteria applied to the Berlind's sample, we have identified G and NG groups in the Millennium group/cluster sample at different redshifts ($z=0,0.5,1.0,1.5$). We split the sample in four mass intervals to probe the approach to equilibrium as a function of mass.\footnote{The average mass for systems with $\sim 20$ galaxies in the Millennium sample is ${\rm M\simeq 10^{13.75}~M_\odot}$.} Our results are presented in Figure 5. We note that even for the very massive systems (${\rm M> 10^{14.25}~M_\odot}$)
there is still an $\sim15\%$ of NG systems. This may serve as a cautionary word for using the virial equation to estimate mass - velocity dispersion may be overestimated for the NG systems. This percentage of NG groups is essentially constant up to $z=1.0$, dropping to around 45\% at redshift 1.5. It is interesting to note that the number of G systems decreases with redshift, regardless of the mass of the system, which is what we should expect in a hierarchical scheme of structure formation. At $z=0$
we find total fraction of G groups in the Millenium sample to be 63\% (orange line), in agreement to what we have found for all groups in the Berlind sample (67\%). This result reinforce the usage of the HD parameter to set the gaussianity of galactic systems, for it is the first time this quantity is used in the astronomical context. 

We also should note that the fraction of G systems for groups/clusters with ${\rm M\geq 10^{13.75}~M_\odot}$ ($\sim$80\%) is significantly higher than that found for systems with ${\rm M< 10^{13.75}~M_\odot}$ ($\sim$45\%). This systematic goes up to redshift 1.0, when then there is a significant drop in the number of G massive systems. For the Berlind sample we find that these fractions for systems with ${\rm M\leq 10^{13.75}~M_\odot}$ and ${\rm M> 10^{13.75}~M_\odot}$ are 76\% and 54\%, respectively, fully consistent to what we find in the Millennium sample. 

\begin{figure} 
\begin{center}
\includegraphics[scale=0.4]{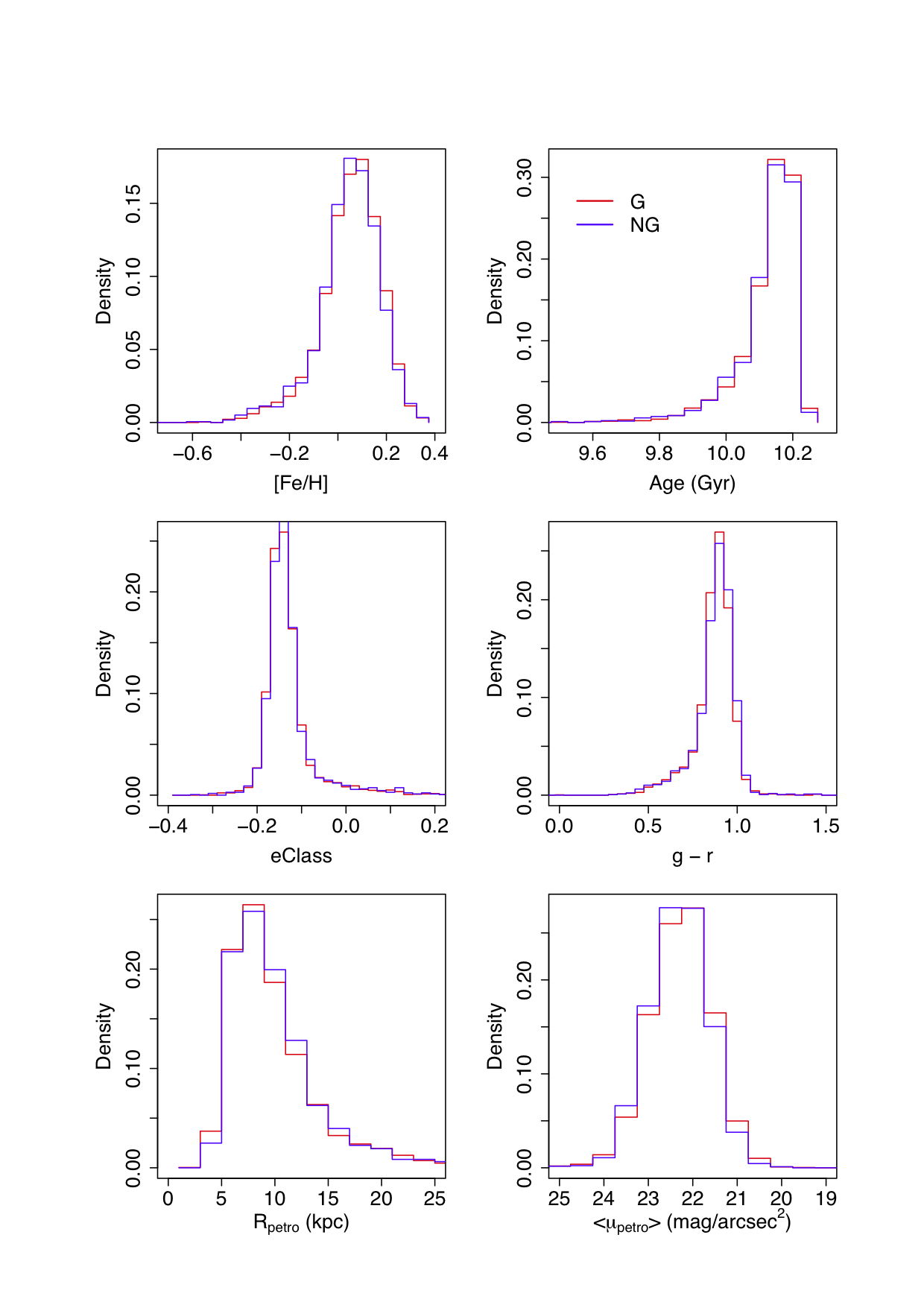}
\caption{Comparison of galaxy properties between G (red lines) and NG (blue and slightly thicker lines) groups for the {\it bright inner} sample $\mathcal{M}_r < -20.7$ \& $R_{\rm norm} < R_{\rm median} = 0.5 R_{\rm 200}$.}
\end{center}
\label{}
\end{figure}

\begin{figure} 
\begin{center}
\includegraphics[scale=0.4]{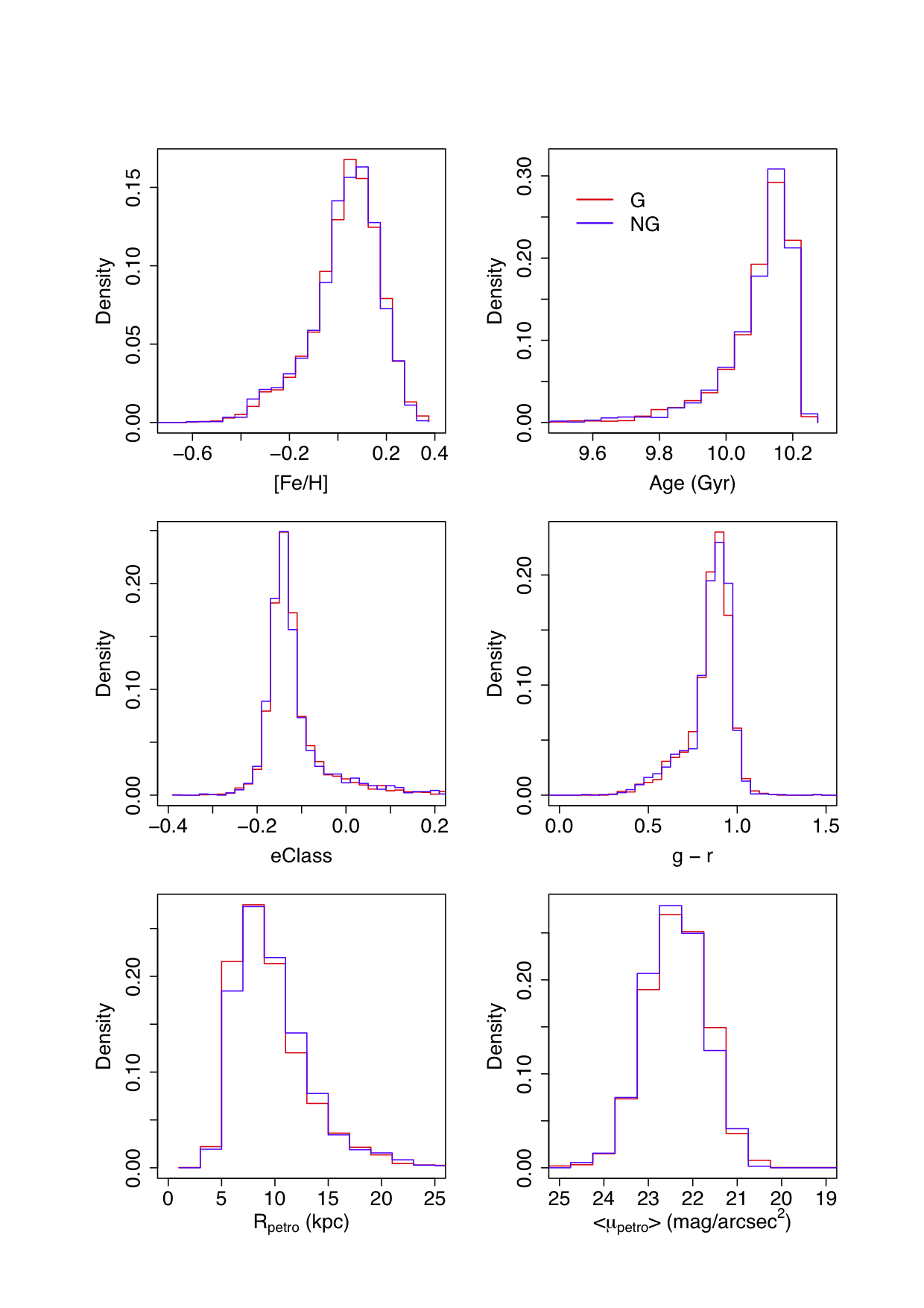}
\caption{Comparison of galaxy properties between G (red lines) and NG (blue and slightly thicker lines) groups for the {\it bright outer} sample $\mathcal{M}_r < -20.7$ \& $R_{\rm norm} > R_{\rm median} = 0.5 R_{\rm 200}$.}
\end{center}
\label{}
\end{figure}

\begin{figure} 
\begin{center}
\includegraphics[scale=0.4]{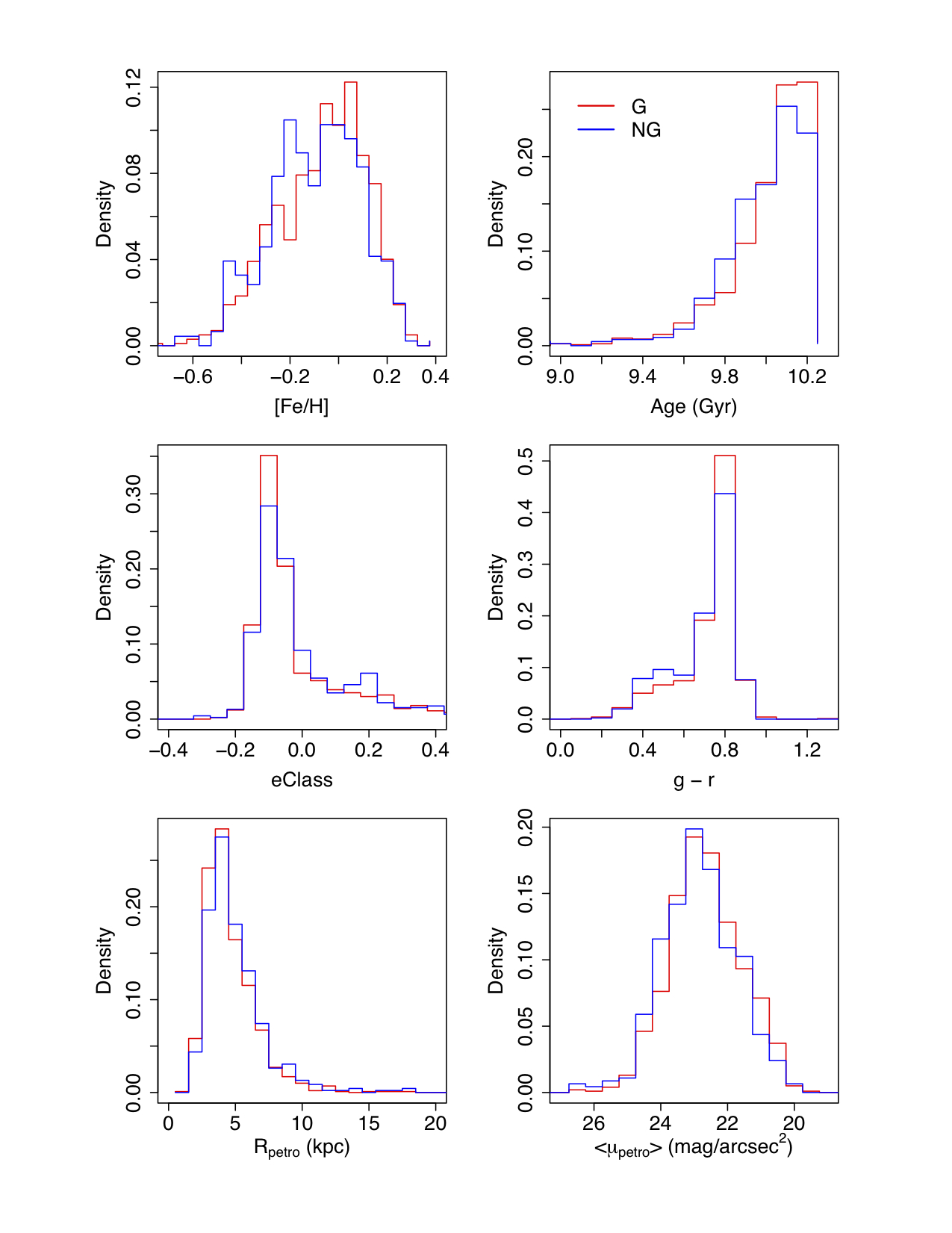}
\caption{Comparison of galaxy properties between G (red lines) and NG (blue slightly thicker lines) groups for the {\it faint inner} sample $-17.92 > \mathcal{M}_r \geq -20.7$ \& $R_{\rm norm} < R_{\rm median} = 0.5 R_{\rm 200}$.}
\end{center}
\label{}
\end{figure}

\begin{figure} 
\begin{center}
\includegraphics[scale=0.4]{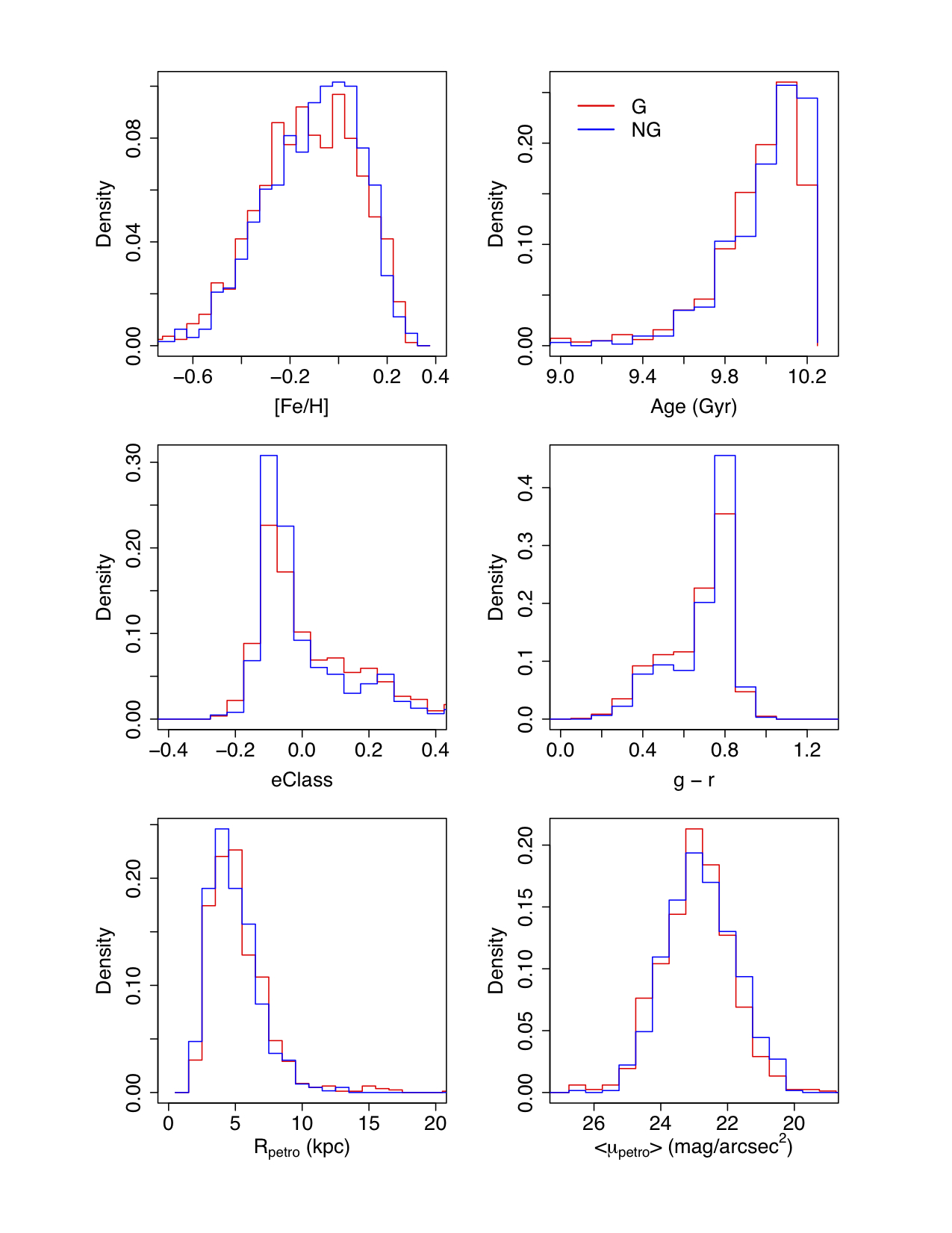}
\caption{Comparison of galaxy properties between G (red lines) and NG (blue slightly thicker lines) groups for the {\it faint outer} sample $-17.92 > \mathcal{M}_r \geq -20.7$ \& $R_{\rm norm} > R_{\rm median} = 0.5 R_{\rm 200}$.}
\end{center}
\label{}
\end{figure}

\section{Properties of galaxies in G and NG groups}

Here in this section, we investigate a possible connection between galaxy evolution and the dynamical state of groups. We analyzed the properties of galaxies belonging to each type of group, namely G and NG. The basic idea is to establish if the dynamical status of a group determines the characteristics of a galaxy residing in it. This is a longstanding problem with no obvious answer (e.g. Biviano et al. 2002, Ribeiro et al. 2010). Taking advantage of the updated FoF group catalog from the SDSS-DR7 (Berlind et al. 2006), for each galaxy we have collected structural parameters (the Petrosian radius - R$_{\rm Petro}$, and the mean surface brightness within R$_{\rm Petro}$ - $\langle\mu_{\rm Petro}\rangle$) and stellar population parameters (Age and Fe/H), from SDSS-DR7 and Mateus et al. (2005). Besides, we used two other parameters that describe galaxies in a more global sense: the {\it eclass}, a single-parameter classifier based on the first two eigencoefficients obtained from the PCA analysis (Yip et al. 2004); and the color (g-r) estimated inside the Petrosian radius in r-band. Two galaxy samples were considered: the {\it{bright}} sample, constituted by galaxies with M$_{\rm r} <$-20.7 and up to the redshift limit of the entire sample, z $<$ 0.095 and the {\it{faint}} sample, composed by galaxies with -17.9 $>$M$_{\rm r} \geq$-20.7 and up to redshift 0.03. In both cases, we keep the spectroscopic magnitude limit of 17.77 in r band. In order to avoid incompleteness at the lower mass regime, only groups with log M $>$13.75 were selected for this study. For each of these two samples we computed the median of radial distances, normalized to the virial radius, and examined the properties of galaxies accordingly to their normalized radial distance R: the {\it inner} sample, R $<$ R$_{\rm median}$ and the {\it outer} sample. R $>$ R$_{\rm median}$, where R$_{\rm median}$ is the median radial distance of the galaxies in both samples.
Figures 5 to 8 display the distributions of the quantities describing the galaxy properties, discriminating between G (red lines) and NG (blue lines) groups. Figures 5 and 6 refer to the {\it bright}, inner and outer samples respectively, and Figures 7 and 8 to the {\it faint}, inner and outer samples. Tables 3 and 4 list the median, the skewness and kurtosis for each of the parameters examined here.

From Figures 6 and 7, we see that for the bright galaxies there are no striking differences between the G and NG group galaxies, a fact that is corroborated by the measured p-values of the KS test shown in Table 3.  When comparing the G and NG group galaxies, only g-r and R$_{\rm petro}$ of the inner sample galaxies display significantly different distributions, whereas for the outer sample only the distribution of R$_{\rm petro}$ show significant differences. As it is apparent from these figures and confirmed by the skewness and kurtosis values in Table 3, almost all distributions deviate significantly from normal except for  $\langle\mu_{\rm Petro}\rangle$ and mainly when considering the outer sample galaxies.

A completely different scenario appears when we look for the galaxies in the {\it faint} samples. From Figures 8 and 9 it is visually apparent how different the distributions are. Results from the KS test, in Table 4, indicate that all distributions are different at confidence levels $\gtrsim$95\%, with a possible exception for the distribution of R$_{\rm petro}$ for galaxies in the outskirts of groups. These results indicate that faint galaxies in G groups exhibit quite different properties when compared to their equivalents in NG groups.

An even stronger result is shown in Table 5 where we examine the differences between the galaxy properties for the inner and outer samples of galaxies belonging to the G and NG groups. Regardless
in what kind of group they are (G or NG ones), the bright galaxies do not display differences in their stellar population parameters (Age and metallicity), but show significant differences in their structural parameters (R$_{\rm Petro}$ and $\mu_{\rm Petro}$) in the sense that bright galaxies tend to be smaller and more concentrated when located in the inner parts of groups. A similar and even stronger effect is also seen for the faint galaxies belonging to G groups. The striking difference is seen when we examine the faint galaxies belonging to NG groups. In this case, the inner and outer galaxies do not show significant differences in any of their properties, suggesting that NG groups faint galaxies are not spatially segregated by their intrinsic properties.


\begin{center}
 \begin{table}
 \caption{Median values, skewness, kurtosis, and the
KS p-value of the distribution of galaxy properties for the subsample
at z $< 0.095$.}           
 \label{}   
\small{   
 \begin{tabular}{p{8mm}p{8mm}cp{10mm}p{10mm}p{10mm}}       
\hline  
\noalign{\smallskip}
\multicolumn{6}{c}{{\large $\mathcal{M}_r < -20.7$ \& $R_{\rm norm} < R_{\rm median}$}}   \\
\noalign{\smallskip}
\hline
           &                &  Median  &  Skewness  &  Kurtosis & ~~~KS  \\
\hline \hline 
 [Fe/H]    &                &                     &         &        &  0.07952   \\
           &             G  &   0.039 $\pm$ 0.007 &  -0.83  &  4.31  &            \\
           &            NG  &   0.034 $\pm$ 0.010 &  -1.24  &  7.93  &            \\
      Age  &                &                     &         &        &  0.62335   \\
           &             G  &   10.13 $\pm$ 0.04  &  -2.19  & 10.30  &            \\
           &            NG  &   10.13 $\pm$ 0.05  &  -2.44  & 12.53  &            \\
   eClass  &                &                     &         &        &  0.12655   \\
           &             G  &  -0.152 $\pm$ 0.002 &   2.82  & 15.34  &            \\
           &            NG  &  -0.150 $\pm$ 0.002 &   2.98  & 16.73  &            \\
  (g - r)  &                &                     &         &        &  0.00009   \\
           &             G  &   0.861 $\pm$ 0.001 &  -1.32  &  6.15  &            \\
           &            NG  &   0.871 $\pm$ 0.001 &  -1.36  &  6.14  &            \\
$R_{\rm Petro}$   &         &                     &         &        &  0.02822   \\
           &             G  &   7.67 $\pm$ 0.01   &   1.38  &  4.99  &            \\
           &            NG  &   7.99 $\pm$ 0.01   &   1.38  &  5.18  &            \\
$\langle\mu_{\rm Petro}\rangle$  &  &             &         &        &  0.05997   \\
           &             G  &  21.993 $\pm$ 0.002 &   0.34  &  3.95  &            \\
           &            NG  &  22.049 $\pm$ 0.003 &   0.20  &  3.24  &            \\
\hline \hline
\noalign{\smallskip}
\multicolumn{6}{c}{{\large $\mathcal{M}_r < -20.7$ \& $R_{\rm norm} > R_{\rm median}$}}   \\
\noalign{\smallskip}
\hline 
           &                &  Median &  Skewness  &  Kurtosis & KS  \\
\hline  \hline
   [Fe/H]  &                &                      &         &        &  0.77019   \\
           &             G  &   0.024 $\pm$ 0.007  &  -0.76  &  3.80  &            \\
           &            NG  &   0.022 $\pm$ 0.009  &  -0.83  &  3.79  &            \\
      Age  &                &                      &         &        &  0.89334   \\
           &             G  &  10.10 $\pm$ 0.04    &  -2.25  & 11.25  &            \\
           &            NG  &  10.11 $\pm$ 0.05    &  -2.27  & 10.61  &            \\
   eClass  &                &                      &         &        &  0.67763   \\
           &             G  &  -0.145 $\pm$ 0.002  &   2.40  & 11.00  &            \\
           &            NG  &  -0.146 $\pm$ 0.002  &   2.31  & 10.19  &            \\
  (g - r)  &                &                      &         &        &  0.26233   \\
           &             G  &   0.846 $\pm$ 0.001  &  -1.27  &  5.04  &            \\
           &            NG  &   0.850 $\pm$ 0.001  &  -1.33  &  5.05  &            \\
$R_{\rm Petro}$   &         &                      &         &        &  0.01502   \\
           &             G  &   7.88 $\pm$ 0.01    &   1.27  &  5.07  &            \\
           &            NG  &   8.17 $\pm$ 0.01    &   1.22  &  4.85  &            \\
$\langle\mu_{\rm Petro}\rangle$  &  &              &         &        &  0.11837   \\
           &             G  &  22.102 $\pm$ 0.002  &   0.18  &  3.57  &            \\
           &            NG  &  22.143 $\pm$ 0.003  &   0.14  &  3.12  &            \\
\hline
 \end{tabular}
}
 \end{table}
 \end{center}


 \begin{center}
 \begin{table}
  \caption{Median values, skewness, kurtosis and the KS p-value of the distribution of galaxy properties for the subsample at z $< 0.03$.}           
  \label{}   
 \small{   
  \begin{tabular}{p{8mm}p{8mm}cp{10mm}p{10mm}p{10mm}}  
 \hline  
\noalign{\smallskip}
 \multicolumn{6}{c}{{\large $-17.92 > \mathcal{M}_r \geq -20.7$ \& $R_{\rm norm} < R_{\rm median}$}}   \\
\noalign{\smallskip}
 \hline
            &                &    Median  &  Skewness  &  Kurtosis & ~~~KS \\
 \hline \hline
   [Fe/H]  &                &                    &         &        &  0.01076   \\
           &             G  &  -0.07 $\pm$ 0.01  &  -0.65  &  3.73  &            \\
           &            NG  &  -0.11 $\pm$ 0.02  &  -0.56  &  3.92  &            \\
   Age     &                &                    &         &        &  0.00228   \\                                                                                                
           &             G  &  10.03 $\pm$ 0.06  &  -1.59  &  5.90  &            \\
           &            NG  &   9.99 $\pm$ 0.09  &  -1.34  &  5.46  &            \\
  eClass   &                &                    &         &        &  0.02216   \\      
           &             G  & -0.100 $\pm$ 0.003 &   1.53  &  4.62  &            \\
           &            NG  & -0.087 $\pm$ 0.005 &   1.23  &  3.94  &            \\
  (g - r)  &                &                    &         &        &  0.00871   \\      
           &             G  &  0.719 $\pm$ 0.001 &  -1.30  &  4.40  &            \\
           &            NG  &  0.702 $\pm$ 0.002 &  -1.01  &  3.03  &            \\
  $R_{\rm Petro}$  &        &                    &         &        &  0.01934   \\     
           &             G  &   3.65 $\pm$ 0.01  &   1.91  &  9.82  &            \\
           &            NG  &   3.96 $\pm$ 0.02  &   2.63  & 14.63  &            \\
  $\langle\mu_{\rm Petro}\rangle$ &  &           &         &        &  0.09507   \\     
           &             G  &  22.46 $\pm$ 0.04  &  -0.09  &  2.74  &            \\
           &            NG  &  22.62 $\pm$ 0.06  &  -0.04  &  2.79  &            \\
 \hline \hline
\noalign{\smallskip}
\multicolumn{6}{c}{{\large $-17.92 > \mathcal{M}_r \geq -20.7$ \& $R_{\rm norm} > R_{\rm median}$}}   \\
\noalign{\smallskip}
\hline 
           &                &    Median  &  Skewness  &  Kurtosis & KS  \\
\hline  \hline
   [Fe/H]  &                &                     &         &        &  0.02939   \\
           &             G  &  -0.15 $\pm$ 0.01   &  -0.51  &  3.20  &            \\
           &            NG  &  -0.11 $\pm$ 0.02   &  -0.57  &  3.13  &            \\
   Age     &                &                     &         &        &  0.00004   \\                                                                                                
           &             G  &   9.96 $\pm$ 0.07   &  -1.38  &  5.34  &            \\
           &            NG  &  10.01 $\pm$ 0.08   &  -1.28  &  4.69  &            \\
  eClass   &                &                     &         &        &  0.00051   \\      
           &             G  &  -0.064 $\pm$ 0.004 &   0.84  &  2.81  &            \\
           &            NG  &  -0.086 $\pm$ 0.004 &   1.21  &  3.57  &            \\
  (g - r)  &                &                     &         &        &  0.00010   \\      
           &             G  &   0.679 $\pm$ 0.001 &  -0.74  &  2.54  &            \\
           &            NG  &   0.704 $\pm$ 0.002 &  -1.03  &  3.08  &            \\
  $R_{\rm Petro}$  &        &                     &         &        &  0.05538   \\     
           &             G  &   4.30 $\pm$ 0.02   &   2.09  & 10.68  &            \\
           &            NG  &   4.07 $\pm$ 0.02   &   1.13  &  5.00  &            \\
  $\langle\mu_{\rm Petro}\rangle$ &  &            &         &        &  0.18048   \\     
           &             G  &  22.67 $\pm$ 0.005  &  -0.08  &  3.19  &            \\
           &            NG  &  22.60 $\pm$ 0.005  &  -0.10  &  2.65  &            \\
\hline
 \end{tabular}
}
 \end{table}
 \end{center}

\normalsize

\begin{center}
\begin{table}
 \caption{Difference between the properties of central and outskirt galaxies.}
 \begin{tabular}{lrr}

 \hline
\noalign{\smallskip}
 \multicolumn{3}{l}{{\large {z $<0.095$ \& $\mathcal{M}_r < -20.7$}}} \\
\noalign{\smallskip}
\hline
                                        &    \multicolumn{1}{c}{G}   & \multicolumn{1}{c}{NG}  \\
 \hline\hline                                        
 
 $\Delta$[Fe/H]                         &   $  0.01  \pm  0.01 $   &   $ 0.01 \pm  0.01 $   \\
  
 $\Delta$Age                            &   $  0.02  \pm  0.05 $   &   $ 0.02 \pm  0.07  $   \\
  
 $\Delta$eClass                         &   $-0.007  \pm  0.002$   &   $-0.004 \pm  0.003$   \\
  
 $\Delta$(g - r)                        &   $ 0.015  \pm  0.001$   &   $ 0.022 \pm  0.001$   \\
  
 $\Delta R_{\rm Petro}$                 &   $-0.21   \pm  0.01 $   &   $-0.18  \pm  0.01 $   \\
  
 $\Delta\langle\mu_{\rm Petro}\rangle$  &   $-0.108  \pm  0.003$   &   $-0.094  \pm  0.004$   \\
 
  &  &  \\
 \hline
\noalign{\smallskip}
 \multicolumn{3}{l}{{\large {z $< 0.03$ \& $-17.92 > \mathcal{M}_r > -20.7$}}} \\
\noalign{\smallskip}
\hline
                                        &    \multicolumn{1}{c}{G}   & \multicolumn{1}{c}{NG}  \\
 \hline\hline                                        
 
 $\Delta$[Fe/H]                         &   $  0.08  \pm  0.02 $   &   $ 0.003  \pm  0.02 $   \\
  
 $\Delta$Age                            &   $  0.07  \pm  0.09 $   &   $ -0.02  \pm  0.12  $   \\
  
 $\Delta$eClass                         &   $-0.035  \pm  0.005$   &   $-0.001  \pm  0.006$   \\
  
 $\Delta$(g - r)                        &   $ 0.041  \pm  0.002$   &   $-0.002  \pm  0.003$   \\
  
 $\Delta R_{\rm Petro}$                 &   $-0.65   \pm  0.02 $   &   $ -0.11  \pm  0.03 $   \\
  
 $\Delta\langle\mu_{\rm Petro}\rangle$  &   $-0.213  \pm  0.006$   &   $ 0.019  \pm  0.008$   \\
 \hline
 \end{tabular}

\end{table}
\end{center}

\section{Discussion}

Several works have shown that the analysis of velocity distributions can separate galaxy groups in relaxed or non-relaxed systems (e.g. Hou et al. 2009; Ribeiro et al. 2011; Einasto et al. 2012ab, Krause et al. 2013). Actually, several important properties of groups can be studied from this perspective. For instance, Hou et al. (2009,2012) found rising velocity dispersion profiles for NG groups; they also have shown that the majority of NG groups have substructures. Ribeiro et al. (2010) found that galaxies are redder and brighter in G groups out to 4${\rm R_{200}}$. Martinez \& Zandivarez (2011) have studied the luminosity functions of G and NG groups, concluding that G groups have a brighter characteristic magnitude. What all these works have in common is the application of normality tests to classify groups as G and NG systems. Clearly, different results can be obtained according to the choice of the specific statistical test used. 

In the present work, we have discussed the dependence of normality tests on the nature of non-Gaussianity and the sample size. Also, we have proposed that normality departures can be reliably quantified by the use of the Hellinger Distance -- see Section 3.3. After a comparison of methods, we verified the superiority of the HD over some of the most used statistical tests (Anderson-Darling, Shapiro-Wilks, and Robust Jarque-Bera). The HD measurement produces almost no type I error, and makes fewer type II errors than the usual normality tests found in the astronomical literature. We also verified, however, that normality departures found by the HD for multimodal systems can be insufficient to reject them as G systems. This has lead us to employ the Dip test (for poorer systems: ${\rm N\leq 20}$) and the Mclust clustering technique (for richer systems: ${\rm N> 20}$) to better detect multimodality -- see Section 3.4. 

We have introduced two criteria to define a relaxed system: the velocity distribution 
should admit just a small deviation from normality (HD $<$ 0.05 after the normalization made in Section 3.3) and it should be unimodal (according to the Dip test at 95\% confidence level, and the Mclust diagnostics). Applying these criteria to the Berlind sample we find that 67\% of the groups are G (with 66\% in poorer, and 70\% in richer systems). We also find that modes (subunits) in multimodal groups are significantly more massive in richer systems  than in poorer ones, $\langle M_{N>20}^{modes}\rangle \simeq (14\pm 2) \langle M_{N\leq 20}^{modes}\rangle$, indicating that larger systems accreet more massive subunits over the process of structure formation. At the same time, we find less low multiplicity NG groups with substructures (6-19\%) than expected (34\%), and that high multiplicity NG groups have a larger amount of substructures (a result similar to Einasto et al. 2012ab). This situation may also derive from the lower power of substructure tests for poor systems (see e.g. Hou et al. 2012).  This could also explain why (for low multiplicity systems) no method indicates that NG groups have higher fractions of substructures than G groups. So, one can just set a lower limit for the fraction of substructures in systems with $N\leq 20$.  The criteria $\mathcal{C}1$ and $\mathcal{C}2$ defined in Section 4 seem more reliable (less prone to type I and II errors) to identify non-relaxed systems than substructure tests. With this in mind, results for the Millennium sample may provide trustworthy virialization rates for galaxy systems with $z\leq 1.5$. In this respect, we verified that these rates strongly depend on the cluster mass. At $z=0$, systems with ${\rm M> 10^{13.75}~M_\odot}$ are $\sim$80\% virialized, whereas systems with ${\rm M\leq 10^{13.75}~M_\odot}$ are only $\sim$45\% virialized. 

Our results suggest that more massive sytems at the present epoch may have grown faster through the hierarchy and now are less affected by the infalling matter, thus preserving their dynamical equilibrium. On the contrary, less massive objects may have formed in poorer regions, thus evolving more slowly through the hierarchy and having their equilibrium more affected by tidal effects and accretion from their surroundings. We see in Figure 5 the increasing virialization rates for groups in all ranges of mass. Note that the curve for systems with ${\rm M\leq 10^{13.75}~M_\odot}$ has the less steep behavior in comparison to the others. This might indicate that galaxy groups experience a strong evolution since $z=1.5$, with their complex mass assembly history ultimately driving a significant fraction of them to the virial condition. Evolution seems to happen more intensely for more massive systems, both providing and preserving equilibrium. As such, our work points to an increasing virialization rate both with time and mass, which is fairly consistent with the results from Araya-Melo et al. (2009), who find that 68-83\% of the halos virialized at $z=0$ for ${\rm \Lambda}$CDM models.

Finally, we probe the connection between galaxy evolution and the dynamical state of galaxy groups.
We compare the difference between the properties of central and outskirt galaxies in G and NG groups,
namely the mean differences listed in columns (1) e (2) of Table 5. 
The bright galaxies (with M$_{\rm r} <$-20.7) do not show significant differences in 
the stellar population properties ([Fe/H], Age, eClass, color)
when located either in the inner or outer parts of the groups. Instead, faint galaxies (-17.9 $>$M$_{\rm r} >$-20.7) show a remarkable difference in all quantities, when we pair inner and outer regions of G systems, while NG groups show no segregation in the properties of inner and outer galaxies. This result reinforces the idea of a less intense galaxy evolution in NG groups (e.g. Ribeiro et al. 2010, 2011).

The fact that in G groups we see a clear difference in properties of galaxies located in the inner and outer regions is much consistent with that observed in rich clusters of galaxies by Strom \& Strom (1978); see also Trujillo (2002). It may be attributed to the tidal stripping of galaxies which, by changing the whole structure of galaxies, produces their shrinkage and central brightening (Aguilar \& White, 1985, 1986). Tidal stripping effects are more important the higher the local density of galaxies, suggesting the G groups have higher central densities and probably more dynamically evolved, since  the sense of the evolution of a gravitational collapse is towards higher central densities. Since it is a theoretical expectation that equilibria of gravitational systems is characterized by a (nearly) normal distribution, our results are suggesting that the G groups of our sample have already attained their equilibrium stage. In contrast by showing that galaxies in NG groups are much less affected by the environment, with much less spatial segregation of their properties than in the G groups, our results reinforce the suggestion of a less intense galaxy evolution in NG groups. These groups may be less dense in their inner parts, consistent with the idea that they may still be far from equilibria.

Finally, in addition to all points referred in this work, we must emphasise that classifying galaxy groups 
according to their velocity distribution is quite sensitive to their multiplicity.  The present work
provides a progress on splitting G and NG groups, but further efforts are needed to improve
the statistical approach, especially regarding the low multiplicity groups for which  the methods presented in Section 3
yield less conclusive results.

\section*{Acknowledgments} 
We thank M. Einasto for important suggestions.
We also thank B. Carvalho for helpful discussions on statistics. ALBR thanks for the support of CNPq, grants 306870/2010-0 and 478753/2010-1. PAAL thanks for the support of FAPERJ, process 110.237/2010. 
MT acknowledges the support of FAPESP, process no. 2012/05142-5.


\begin{thebibliography}{99}

\bibitem[]{} Adami, Biviano, \&Mazure(1998)]{ada98} Adami, C.,  Biviano, A., Mazure, A., 1998, A\&A, 331, 493
\bibitem[]{} Aguilar, L. A., \& White, S. D. M. 1985, ApJ, 295, 374
\bibitem[]{} Aguilar, L. A., \& White, S. D. M. 1986, ApJ, 307, 97
\bibitem[]{} Amari, S., 1985,  Differential-geometrical methods in statistics, Lecture Notes in Statistics, vol. 28, Springer 
\bibitem[]{} Amari, S. and Nagaoka, H., 2000, in Methods of Information Geometry, vol. 191, American Mathematical Society and Oxford University Press 
\bibitem[]{} Ashman, L.M., Bird, C.M. \& Zepf, S.E., 1994, AJ, 108, 2348
\bibitem[]{} Araya-Melo, P. A., Reisenegger, A.; Meza, A., van de Weygaert, R., D\"unner, R. \& Quintana, H., 2009, MNRAS, 399, 97
\bibitem[]{} Barnes, E.I,  Williams, L.L.R., 2012, ApJ 748, 144
\bibitem[]{} Beers, T.C., Flynn, K. \& Gebahrdt, K., 1990, AJ, 100, 32
\bibitem[]{} Beraldo e Silva, L.J., Lima, M., Sodr\'e Jr., L., 2013, MNRAS submitted (arXiv:1301.1684v1)
\bibitem[Berlind et al.(2006)]{ber06} Berlind, A.A., Frieman, J., Weinberg, D.H., et al., 2006, ApJS, 167, 1
\bibitem[]{} Biviano, A., Katgert, P., Thomas, T., Adami, C., 2002, A\&A, 387, 8
\bibitem[Biviano et al.(2006)]{biv06} Biviano, A., Murante, G., Borgani, S., Diaferio, A., Dolag, K., Girardi, M., 2006, A\&A, 456, 23
\bibitem[Bruzual \& Charlot(2003)]{BC:03} Bruzual G., Charlot S., 2003, MNRAS, 344, 1000
\bibitem[]{} Carlberg, R. G., Yee, H. K. C. \& Ellingson, E., 1997, ApJ, 478, 462 
\bibitem[]{} Cattaneo, A., Mamon, G.A., Warnick, K. \& Knebe, A., 2011, A\&A, 533, id.A5
\bibitem[Chabrier(2003)]{Chabrier:2003} Chabrier G., 2003, PASP, 115, 763 
\bibitem[Croton, et al.(2006)]{Croton:06} Croton, D., et al. 2006, MNRAS, 365, 11
\bibitem[]{} da Silva, A.F., 2009, Comput. Methods Programs Biomed., 94, 1
\bibitem[]{} de Lucia, G., Springel, V., White, S. D. M., Croton, D. \& Kauffmann, G., 2006, MNRAS, 366, 499
\bibitem[de Lucia \& Blaizot(2007)]{deLB:07} de Lucia, G., Blaizot, J., 2007, MNRAS, 375, 2
\bibitem[]{} Diaferio, A., Ramella, M., Geller, M. \& Ferrari, A., 1993, AJ, 105, 2035
\bibitem[]{} Diebolt, J. \& Robert, C.P., 1994, Journal of the Royal, Series B, 56, 363 
\bibitem[]{} Dressler, A. \& Shectman, S.A., 1988, AJ, 95, 985
\bibitem[]{} Fadda D., Girardi M., Giuricin G., et al., 1996, ApJ, 473, 670
\bibitem[]{} Einasto, M., Tago, E., Saar, E., Nurmi, P., Enkvist, I., Einasto, P., Hein\"am\"aki, P., Liivam\"agi, L. J., Tempel, E., Einasto, J., Mart\'{\i}nez, V. J., Vennik, J. \& Pihajoki, P., 2010, 522, id.A92
\bibitem[]{} Einasto, M., Vennik, J., Nurmi, P., Tempel, E., Ahvensalmi A., Tago, E.,Liivam\"agi, L. J.,  Saar, E.,  
Hein\"am\"aki, P., Einasto, J.,  \& Martínez, V. J., 2012, A\& A, 540, id.A123 (2012a)
\bibitem[]{} Einasto, M., Liivam\"agi, L. J., Tempel, E., Saar, E., Vennik, J., Nurmi, P., Gramann, M., Einasto, J., Tago, E., Hein\"am\"aki, P., Ahvensalmi, A. \& Martínez, V. J., 2012, A\& A, 542, id.A36 (2012b)
\bibitem[]{} Fouqu\'e, P., Gourgoulhon, P., Chamaraux, P. \& Paturel, G., 1992, A\& AS, 93, 211
\bibitem[]{} Fraley, C \& Raftery, A.E., 2007, Journal of Classification, 24,155
\bibitem[Girardi et al.(1998)]{gir98} Girardi, M., Giuricin, G., Mardirossian, F., Mezzetti, M., \& Boschin, W. 1998, ApJ, 505, 74
\bibitem[]{} Giuricin, G., Gondolo, P., Mardirossian, F., Mezzetti, M. \& Ramella, M., 1988, A\&A, 199, 85
\bibitem[]{} Gourgoulhon, P., Chamaraux, P. \& Fouqu\'e, P., 1992, A\&A, 255, 69
\bibitem[]{} Gunn, J.E. \& Gott, J.R, 1972, ApJ, 176, 1
\bibitem[]{} Hansen, S.H., Egli, D., Hollenstein, L., Salzmann, C., 2005, New Astron. 10, 379
\bibitem[]{} Hansen, S.H., Moore, B., Zemp, M., Stadel, J., 2006, JCAP, 1, 14
\bibitem[]{} Hartigan, J. A. \&  Hartigan, P.M., 1985, The Annals of Statistics, 13, 70
\bibitem[]{} Hjorth, J., Williams, L.L.R., 2010, ApJ, 722, 851
\bibitem[]{} Hou, A., Parker, L., Harris, W. \& Wilman, D.J., 2009, ApJ, 702, 1199
\bibitem[]{} Hou, A., Parker, L. C., Wilman, D.J., McGee, S. L., Harris, W. E., Connelly, J. L., Balogh, M. L., Mulchaey, J. S. \& Bower, Richard G., 2012, MNRAS, 421, 3594
\bibitem[]{} Jarque, C. M. \& Bera, A. K., 1987, International Statistical Review 55, 163
\bibitem[]{} Kass, R. and Vos, P., 1997,  Geometrical Foundations of Asymptotic Inference, Wiley Series in Probability and Statistics, John Wiley and Sons, NY, USA
\bibitem[]{} King, I.R., 1966, AJ, 71, 64
\bibitem[]{} Kolmogorov, A.N., 1933,  Foundations of the Probability Theory, Chelsea, New York.
\bibitem[La Barbera et al.(2010)]{lab10} La Barbera, F., Lopes, P.A.A., de Carvalho, R. R., de La Rosa, I. G., Berlind, A. A., 2010, MNRAS, 408, 1361
\bibitem{}{} Krause, M.O., Ribeiro, A.L.B. \& Lopes, P.A.A., 2013, A\&A, 551, id.A143
\bibitem[]{} Langmuir, I., 1925, Phys. Rev., 26, 585
\bibitem[]{} LeCam, L.M., 1986,  Asymptotic Methods in Statistical Decision Theory, Springer-Verlag, New York.
\bibitem[Lemson et al.(2006)]{Lemson:2006} Lemson, G., et al. (The Virgo Consortium), 2006, preprint (astro-ph/0608019)
\bibitem[]{} Lemze, D., Wagner, R., Rephaeli, Y., Sadeh, S., Norman, M.L., Barkana, R., Broadhurst T., Ford, H., Postman, M., 2012, ApJ, 752, 141
\bibitem[\protect\citeauthoryear{Lopes}{2007}]{lop07} Lopes, P.A.A., 2007, MNRAS, 380, 1680
\bibitem[\protect\citeauthoryear{Lopes et al.}{2009a}]{lop09a} Lopes,  P.A.A., de Carvalho, R.R., Kohl-Moreira, J.L., Jones, C., 2009a, MNRAS, 392, 135
\bibitem[\protect\citeauthoryear{Lopes et al.}{2009b}]{lop09b} Lopes, P.A.A., de Carvalho, R.R., Kohl-Moreira, J.L., Jones, C., 2009b, MNRAS, 399, 2201
\bibitem[]{} Lynden-Bell, D., 1967, MNRAS, 136, 101
\bibitem[]{} Mamon, G., Dynamical Theory of Groups and Clusters of Galaxies, in Gravitational Dynamics and the N-body Problem, ed. by F. Combes, E. Athanassoula, 1993, p. 188.
\bibitem[]{} Mamon, G. in Groups of Galaxies in the Nearby Universe, Proceedings of the ESO Workshop held at Santiago de Chile. Edited by I. Saviane, V.D. Ivanov \& J. Borissova, 2007, p. 203 
\bibitem[]{} Mart\'{\i}nez, H. J.; Zandivarez, A., 2012, MNRAS, 419, L24
\bibitem[]{} Menci, N. \& Fusco-Femiano, R., 1996, ApJ, 472, 46
\bibitem[]{} Merrall, T.E.C., Henriksen, R.N., 2003,  Ap.J., 595, 43
\bibitem[]{} Muratov, A.L. \& Gnedin, O.Y., 2010, ApJ, 718, 1266
\bibitem[]{} Naab, T., Johansson, P., Ostriker, J.P. \& Efstathiou, G., 2007, ApJ, 658, 710
\bibitem{}{} Niemi, S., Nurmi, P., Hein\"am\"aki, P. \& Valtonen, M., 2007, MNRAS, 382, 1864
\bibitem{}{} Nolan, J. P., 1998, Statistics and Probability Letters, 38, 187.
\bibitem[]{} Ogorodnikov, K. F. 1957, Sov.Astron, 1, 748 
\bibitem[]{} Pinkney, J., Roettiger, K., Burns, J.O. \& Bird, C., 1996, ApJ, 104, 1
\bibitem[Popesso et al.(2005)]{pop05} Popesso, P., Biviano, A., B\"ohringer, H., Romaniello, M., Voges, W., 2005, A\&A, 433, 431
\bibitem[\protect\citeauthoryear{Popesso et al.}{2007}]{pop07} Popesso, P., Biviano, A., B\"ohringer, H., Romaniello, M., 2007, A\&A, 464, 451
\bibitem[]{} Robotham, A., Phillipps. S. \& De Propis, R., 2008, ApJ, 672, 834
\bibitem[]{} Ribeiro, A.L.B., Lopes, P.A.A. \& Trevisan, M., 2010, MNRAS, 409, L124
\bibitem[]{} Ribeiro, A.L.B., Lopes, P.A.A. \& Trevisan, M., 2011, MNRAS, 413, L81
\bibitem[]{} Ruckdeschel, Kohl, M., Stabla, T. \& Camphausen, F., 2006, R News, 6, 2
\bibitem[]{} Seier, E., 2011, Normality Tests- Power Comparison -- International Encyclopedia of Statistical Science, Part 14, 1000-1003, Miodrag Lovric (ed), Springer
\bibitem[]{} Shapiro, S. \& Wilk, M. B., 1965, Biometrika 52, 591
\bibitem[]{} Skielboe, A., Wojtak, R., Pedersen, K., Rozo, E., Rykoff, E.S.,  2012, ApjL, 758, 16
\bibitem[]{} Springel, V., White, S.D.M., Tormen, G. \& Kauffmann, G., 2001, MNRAS, 328, 726
\bibitem[]{} Springel, V., et al. 2005, Nature, 435, 629
\bibitem[]{} Strom, S. E., \& Strom, K. M. 1978, AJ, 83, 73
\bibitem[]{} Thadewald, T. and B\"uning, H., 2007, Journal of Applied Statistics 34, 87
\bibitem[]{} Trujillo, I., Aguerri, J. A. L., Guti\'errez, C. M., Caon, N., Cepa, J. 2002, ApJ, 573, L9
\bibitem[]{} West, M., Oemler, A. \& Dekel, A., 1988, ApJ, 327, 1
\bibitem[]{} Yahil, A. \& Vidal, N. V., 1977, ApJ, 214, 347
\bibitem[]{} Yip, C. W., Connolly, A. J., Szalay, A. S., Budav\'ari, T., SubbaRao, M., Frieman, J. A., Nichol, R. C., Hopkins, A. M., York, D. G., Okamura, S., Brinkmann, J., Csabai, I., Thakar, A. R., Fukugita, M., Ivezi\'c, Z., 2004, AJ, 128, 585


\end{thebibliography}
\end{document}